\documentclass[10pt,conference]{IEEEtran}

\input epsf
\usepackage{graphicx}
\usepackage{multirow}
\usepackage{footnote, url}
\usepackage[flushleft]{threeparttable}
\usepackage{balance}
\usepackage{scrextend}
\deffootnote[0.0em]{0.0em}{2em}{\textsuperscript{\thefootnotemark}\,}

\newcommand{\accepted}[1]{}

\newcommand{\heng}[1]{}
\newcommand{\xingfang}[1]{}
\newcommand{\Foutse}[1]{}
\newcommand{\Washi}[1]{}

\usepackage{tcolorbox}
\tcbuselibrary{skins, breakable, theorems}

\newtcbtheorem{answer}{Answer}%
  {enhanced, breakable,
    colback = fbbg, colframe = gray, colbacktitle = fbtitle,
    attach boxed title to top left = {yshift = -2.2mm, xshift = 0mm},
    boxed title style = {rounded corners},
    separator sign none,
    fonttitle = \sffamily\bfseries}{qst}

\begin{document}

\bibliographystyle{unsrt}

\definecolor{fbtitle}{HTML}{636463}
\definecolor{fbbg}{HTML}{F2F2F2}

% \title{\textbf{\Large Identifying and Exploiting Duplicate Technical Forum Posts with GPT-3}}

%\title{\textbf{\Large Refining GPT-3 Embeddings with a Siamese Structure\\ for Technical Post Duplicate Detection}}
\title{Refining GPT-3 Embeddings with a Siamese Structure for Technical Post Duplicate Detection}

\author{\IEEEauthorblockN{Xingfang Wu*, Heng Li*, Nobukazu Yoshioka\dag, Hironori Washizaki\dag, Foutse Khomh*}
\IEEEauthorblockA{
*Polytechnique Montréal, Montréal, Canada, \textit{\{firstname.lastname\}}@polymtl.ca \\
\dag Waseda University, Tokyo, Japan, nobukazuy@acm.org, washizaki@waseda.jp
}}

%+++++++++++++++++++++++++++++++++++++++++++

% use only for invited papers
%\specialpapernotice{(Invited Paper)}

% make the title area
\maketitle

\begin{abstract}
% \secdescription{1. duplicate questions background. 2. Finding duplicate benefits. 3. duplicate mining, one vs one comparison is not scalable to the overwhelming amount of new posts. 4. more advanced ways, various approaches in representing the forum posts. but performance may suffer, as limited amount of domain-specific training data and well-annotated data. Existing works use very old data. duplicate datasets/benchmarks are from years ago. 5. Utilizing LLMs is promising. 6. In this work, we evaluate the off-the-shelf GPT-3 embedding over duplicate detection tasks. We compare the GPT-3 embedding with a benchmark, and the raw embedding can achieve competitive results even without further training. 7. As available datasets are outdated, from years ago. We constructed a duplicate question dataset from Stack Overflow dump (Dec 2022) for better evaluate the embeddings over up-to-date forum posts. 8. To further improve the performance, we design siamese-based network to learn a latent embedding utilizing available annotation. Proved by our experiments, the resulted latent embedding can better capture the duplicate relation of technical forum posts. Achieved Top-X among XX candidate.}
% need to associate with the conference.

One goal of technical online communities is to help developers find the right answer in one place. A single question can be asked in different ways with different wordings, leading to the existence of duplicate posts on technical forums. The question of how to discover and link duplicate posts has garnered the attention of both developer communities and researchers. %Developer communities have developed their own mechanisms to deal with duplication. 
For example, Stack Overflow adopts a voting-based mechanism to mark and close duplicate posts. However, addressing these constantly emerging duplicate posts in a timely manner continues to pose challenges. Therefore, various approaches have been proposed to detect duplicate posts on technical forum posts automatically. The existing methods suffer from limitations either due to their reliance on handcrafted similarity metrics which can not sufficiently capture the semantics of posts\accepted{\heng{we need to have a consistent use of questions vs. posts}}, or their lack of supervision (i.e., leveraging existing duplicate annotations) to improve the performance. Additionally, the efficiency of these methods is hindered by their dependence on pair-wise feature generation, which can be impractical for large amount of data. 
% The existing methods are mainly based on one-versus-one comparisons and adopt various similarity metrics (e.g., topical similarity) to different components of posts.
%\heng{I feel we did not capture the main drawback of the existing approach here: we also use a one-versus-one comparison approach? what's the key difference that contributes to the different efficiency?} 
% However, due to their complex mechanisms, these methods might not scale well with the growing number of technical posts\accepted{\heng{growing number of posts? I feel the scalability is more related to the total number of posts, not the number of duplicates}} and could exhibit limited generalization ability. 
%The emergence of large language models, such as GPT-3, has introduced a new opportunity to tackle this task with embeddings accurately representing semantic similarity.\heng{not clear why GPT-3 can address the mentioned problem (scalability)}. 
In this work, we attempt to employ and refine the GPT-3 embeddings for the duplicate detection task. We assume that the GPT-3 embeddings can accurately represent the semantics of the posts.
% Our experiment on a benchmark dataset shows that semantic embeddings generated by our approach can be used as an effective representation for measuring the similarity of technical posts\accepted{\heng{embeddings represent similarity, not quite accurate... maybe the embeddings ``can be used as an effective representation for measuring the similarity''?}}, which is helpful in detecting duplication. 
In addition, by training a Siamese-based network based on the GPT-3 embeddings, we obtain a latent embedding that accurately captures the duplicate relation in technical forum posts. Our experiment on a benchmark dataset confirms the effectiveness of our approach and demonstrates superior performance compared to baseline methods. When applied to the dataset we constructed with a recent Stack Overflow dump\accepted{\heng{mention the contribution of constructing the dataset first}}, our approach attains a Top-1, Top-5, and Top-30 accuracy of 23.1\%, 43.9\%, and 68.9\%, respectively. With a manual study, we confirm our approach's potential of finding unlabelled duplicates on technical forums. %\Foutse{so what is the finding? it might be better to give a summary of the result!}. 
We released our dataset and code in our supplementary package to promote further studies\footnote{\label{supmat}Supplementary material package:\\\url{https://github.com/mooselab/suppmaterial-PostDupGPT3}}.

% https://github.com/TechnicalForumResearcher/DuplicateDetectionWithGPT3.git

\end{abstract}

% \IEEEoverridecommandlockouts
% \vspace{1.5ex}

\begin{IEEEkeywords}
Technical Forum, Duplicate Detection, Large Language Model, Siamese Network
\end{IEEEkeywords}

% For peer review papers, you can put extra information on the cover
% page as needed:
% \begin{center} \bfseries EDICS Category: 3-BBND \end{center}
%
% for peerreview papers, inserts a page break and creates the second title.
% Will be ignored for other modes.
\IEEEpeerreviewmaketitle

\section{Introduction}\label{sec:intro}

Online developer communities serve as valuable platforms for developers to find solutions to coding problems and seek guidance on implementation choices. The growth of these online communities has been remarkable in recent years, with an increasing number of posts published daily. By July 2023, there were more than 24 million posts on Stack Overflow alone. This large number of technical posts poses challenges to content maintenance and the recommendation systems of these forums. A previous study reveals that approximately 53\% of posts on Stack Overflow were closed as duplicates~\cite{abric2019can}, and the proportion is increasing over time~\cite{correa2013fit}. %Such duplicate questions may seem redundant, as one main goal for these online communities is to help people find answers to their questions rapidly by bringing together correct answers in one place.

% \accepted{\xingfang{modified}\heng{not clear what this sentence mean and its purpose}}

% \xingfang{Problems caused by duplicate questions.}
% \xingfang{the following paragraph seems too much?}
The existence of duplicate posts on technical forums leads to redundancy, and thus poses challenges in managing the forum content. Firstly, duplicate posts may confuse users and increase their efforts when searching for posts relevant to their problems. Secondly, solving duplicate questions may demand extra community support efforts (e.g., moderators' efforts and question repliers' efforts). They may waste time responding to and flagging duplicate posts instead of addressing new or more complex issues. Thirdly, duplicate posts may reduce the engagement of technical communities. For example, domain experts may become less enthusiastic about solving questions in technical communities if they are continually faced with the same or similar questions. Overall, the over-presence of duplicate posts may be harmful to technical communities.

\accepted{\heng{I am not sure about the purpose of this paragraph. Instead, we should discuss the problem of having duplicated questions, for example: 1) confusing users and increasing their efforts when they are searching for posts relevant to their problems; 2) increasing community support efforts (moderators' efforts, question answerers' efforts); 3) if a user posted a duplicate questions, they have wasted time and effort in the first place, as they could have used information in an earlier post to solve their problem.}}

% \xingfang{However, it is beneficial.}

% Although the duplicate questions might waste some time and effort for community members to filter out useful information and answer questions that have already been correctly solved, they are not always detrimental. 

% \heng{This paragraph can be entirely deleted, or briefly mentioned in threats validity, as it does not contribute to the motivation of the work.}
However, duplicates are not always detrimental. Some degree of duplication in technical online communities is even desirable~\cite{so_duplicate}. Technical communities are tailored to a diverse audience, from beginners to programming experts. Duplicate posts with different phrasings enhance coverage, accommodating a wide range of queries. This improves the chances of users finding the answers they seek when these duplicate questions are clearly associated. Furthermore, duplication can lead to more in-depth discussions, additional insights, and new solutions and ideas. Approximately 9\% of duplicates receive more views than their original counterparts despite being closed~\cite{abric2019can}\accepted{\heng{cite the source of the information, or mention ``Based on our analysis''}}. Duplicate questions may also receive distinct answers, which could provide extra value to users~\cite{abric2019can}. Therefore, incorporating and associating duplicate posts may deliver extra value and greater content completeness for users.

Several approaches~\cite{zhang2015multi, ahasanuzzaman2016mining, lau2016empirical, zhang2017feature, zhang2017detecting, zhang2018duplicate, silva2018duplicate} have been proposed to automatically detect duplicate posts of online technical forums.
These approaches primarily tackle this task either by meticulously selecting and extracting textual features from the content of technical posts (e.g., statistic features~\cite{zhang2017feature, zhang2017detecting, zhang2018duplicate} or topic-level features~\cite{zhang2015multi, silva2018duplicate}), or by employing word embeddings trained on large external corpora~\cite{lau2016empirical}. These textual features or embeddings are subsequently used to measure the similarity between each pair of posts, or used as the input of a classifier to determine if the pair is a duplicate or not. 

Nevertheless, there exist two notable challenges in existing approaches that restrict their practical applications.
% \xingfang{modified from here.}
The first challenge is the performance limitations related to the mechanisms of existing methods. Some existing approaches rely on handcrafted features or textual similarity metrics (e.g.,~\cite{zhang2015multi}), which may not have good generalizability. A performance decrease, which may be associated with the variations in dataset utilized in the study, is observed by a reproducibility study for DupPredictor and Dupe~\cite{silva2018duplicate}. %This suggests that they may not exhibit the stable performance across diverse data distributions, such as technical topics and forum types. 
Moreover, previous works utilize embedding techniques (e.g., Doc2Vec~\cite{lau2016empirical} to generate semantic representations for similarity measurement, which miss the opportunity to leverage existing duplicate annotations on online forums (i.e., supervision) to enhance their performance. Semantic embeddings intended for general purposes may not sufficiently capture the intricate nuances in domain-specific technical post data, limiting their effectiveness in identifying duplicate relations within technical posts.\accepted{\heng{Instead of mentioning using a general corpus, it might be better to mention that the embeddings are for general purposes, not for duplication detection purposes, thus to motivate our use of the Siamese structure}}
% Besides, training embedding models may be time-consuming and require large training data.
The second challenge is the scalability issue caused by pair-wise feature generations. Most current approaches (e.g., ~\cite{ahasanuzzaman2016mining, wang2020duplicate}) formulate the duplication detection task as a classification task and require one-versus-one feature generation for post pairs, and thus, are not scalable to the ever-increasing amount of posts. For example, when querying duplicates for a post, Dupe~\cite{ahasanuzzaman2016mining} needs to generate five similarity metrics, including term overlap, entity overlap, etc., for all possible permutation and combination of posts. These pair-wise manipulation can be expensive given the large number of posts and the complexities of calculating similarities for all features.

% \xingfang{end of modification}

In this work, we leverage large language models (LLMs), more specifically, the GPT-3 embeddings, and refine them with a Siamese structure for the purpose of detecting duplicate posts.
%The emergence of Large Language Models (LLMs) provides a new approach for us to represent technical posts and detect duplicates in online communities. 
LLMs, which are typically pre-trained with large general corpora, are expected to capture the semantics of the technical posts in an accurate and generalizable way.\accepted{\heng{do we have a reference for this? if not, we can say: are expected to capture the semantics of the technical posts in an accurate and generalizable way}}
%The off-the-shelf pre-trained models are usually pre-trained with large general corpora and are not limited to specific technical topics. 
%In this work, we introduce a generalizable and efficient approach for detecting duplicates in technical forums by utilizing and refining the GPT-3 embeddings, which, to some extent, address the previously mentioned challenges. 
%Specifically, we leverage a GPT-3 model to generate semantic embeddings for posts and employ a 
The Siamese-based structure~\cite{koch2015siamese}, which consists of two sub-networks sharing the same parameters, takes the GPT-3 embeddings as the input and learns a latent, contrastive embedding from the existing duplicate labels offered by Stack Overflow. In other words, the Siamese-based structure refines the GPT-3 embeddings through the supervision of the duplicate labels. 
%in builting upon these embeddings to acquire a latent, contrastive embedding that more effectively captures post duplicate relationships. 
Our experiments demonstrate that our approach exhibits significantly better accuracy and stability for duplicate post detection across various technical topics. Furthermore, the proposed method, which eliminates the need for pairwise feature generation, offers enhanced scalability.

To understand the effectiveness and characteristics of our proposed approach, we proposed the following research questions (RQs):
% \heng{also need to discuss how you overcome the scalability issue mentioned earlier.}\xingfang{plan to mention somewhere in the following research questions. rather than to set a RQ for scalability?}\heng{no need for an RQ. check earlier suggested flow}

\textbf{RQ1: How does the performance of our proposed approach compare to that of baselines on a benchmark dataset?}

\textbf{RQ2: How does our approach perform on our dataset constructed with all Stack Overflow duplicates? How do training settings (e.g., loss function, batch size) influence the performance?}

\textbf{RQ3: How does the performance of models trained on data specific to a particular topic compare to that of models trained on general data?}

The major contributions of this work are as follows: \accepted{\heng{contributes are not about what we did, but about what we provide/offer (data, approach, insights, lessons learned, etc.): rephrase them}}
\begin{enumerate}
    \item We provide a duplicate question dataset constructed from a recent Stack Overflow dump, consisting of 723,008 duplicate post pairs.
    %\item We propose a scalable approach for detecting duplicate posts on technical forums by utilizing GPT-3 embeddings and existing duplicate annotations. 
    \item We propose a scalable approach for duplicate post detection that leverages GPT-3 embeddings and a Siamese-based structure to refine the embeddings, which outperforms existing baselines or when GPT-3 embeddings are used alone. 
    \item We study the impact of different training configurations (e.g., different loss functions) or scenarios (within a technical topic vs. across technical topics) on the performance of our approach.
    \item Through a manual study of the detection results of our approach, we demonstrate that our approach can detect duplicate posts that have not been but should have been labeled.
    % We formulate duplicate detection as a ranking task, which does not require pair-wise feature generation.
    %\item The latent embeddings obtained by our proposed Siamese-based structure show a superior ability to represent technical posts in terms of their duplicate relations and improve the performance of duplicate detection.
    % \item The latent embeddings obtained by our proposed Siamese-based structure show a superior ability to represent technical posts in terms of their duplicate relations and improve the performance of duplicate detection.
    %\item We broaden our study by assessing our approach across diverse technical topics and validating its ability to identify unlabeled duplicate posts by a manual study.
    % \item We examine how the performance of our approach varies across different topics of training and test data, which enhances the comprehensiveness of our study.
    % \item We elevate the validation and exploration of our approach through a discussion in which we manually analyze the ranking results produced by our method, facilitating a more precise comprehension of its capacity to proficiently identify unlabeled duplicate questions.
    % Our analysis of the duplicate detection results generated by our proposed method offers a deeper\heng{than what?} insight into its potential effectiveness in identifying unlabeled duplicate questions.

\end{enumerate}

% The paper is structured as follows: Section~\ref{sec:background} introduces the background of this study. Section~\ref{sec:related_work} introduces the related works in three domains. Section~\ref{sec:data_construction} elaborates on the process by which we construct our dataset from a recent Stack Overflow dump. In Section~\ref{sec:preliminary}, we conduct a preliminary study on duplicate posts in the Stack Overflow dump. In Section~\ref{sec:method}, we introduce our problem formulation and embedding generation methodology. We also illustrate our proposed Siamese structure and the loss functions we employed in this section. In Section~\ref{sec:eval}, we evaluate our proposed approach with the three research questions. In Section~\ref{sec:discussion}, we conduct a case study to understand our method's potential to detect unlabeled duplicates. Section~\ref{sec:threats} identifies the threats to the validity of our study. At last, we summarize and conclude our work in Section~\ref{sec:conclusions}.

The paper is structured as follows: Section~\ref{sec:background} introduces the study's background, while Section~\ref{sec:related_work} presents related works. % across three related domains. 
Section~\ref{sec:data_construction} details our dataset construction from a recent Stack Overflow dump, followed by a preliminary study on duplicate posts in Section~\ref{sec:preliminary}. Section~\ref{sec:method} provides details of our approach, %including embedding generation, Siamese structure, and loss functions. 
 while Section~\ref{sec:eval} evaluates our approach along three research questions. Section~\ref{sec:discussion} investigates our approach's ability to detect unlabeled duplicates through a case study. Section~\ref{sec:threats} identifies threats to the validity of our study. Section~\ref{sec:conclusions} concludes our work.

% \accepted{\heng{put them as three separate sections: rename ``Motivation'' to ``Background'', Related Work, Dataset Construction and Preliminary Study (the qualitative study).} }

\accepted{\heng{the introduction is too long! We can move some text (the mechanism of Stack Overflow and why duplicates exist) to the background section.}\xingfang{good idea!}}

% \section{Preliminaries} \label{sec:prelim}

\section{Background} \label{sec:background}

\noindent\textbf{Why do duplicates exist in online forums?}
% \xingfang{Why duplicate questions are asked:}
Duplicate questions are very common in online developer communities~\cite{abric2019can}. Naturally, developers may encounter similar or the same questions and formulate them in very different wordings and diverse ways. For example, ``Copy sentence to Clipboard using simple JS'' and ``Copy string to clipboard initiated by click on injected element in JavaScript'' are the titles of a duplicate pair on Stack Overflow\footnote{Question IDs: 34954370 and 43638872.}. In this post pair, authors used different levels of abbreviation for terminologies (i.e., ``JavaScript'' and ``JS''), synonyms (i.e., ``sentence'' and ``string'') and degrees of brevity (i.e., the second title contains more specific requirements). 

The accuracy of the formulated questions on technical forums may vary according to the users’ writing habits, level of expertise and experience in a specific topic. For example, a previous study~\cite{abric2019can} indicates that inexperienced users are more likely to post duplicate questions. This may result from the fact that inexperienced users may have poor ability of question formulation and limited familiarity with topic-specific terminologies. %The higher rate of posting duplicate questions may be associated with their poor posting skills.

\begin{figure}[!ht]
    \centering
    % \vspace{-3mm}
	\includegraphics[scale=0.45]{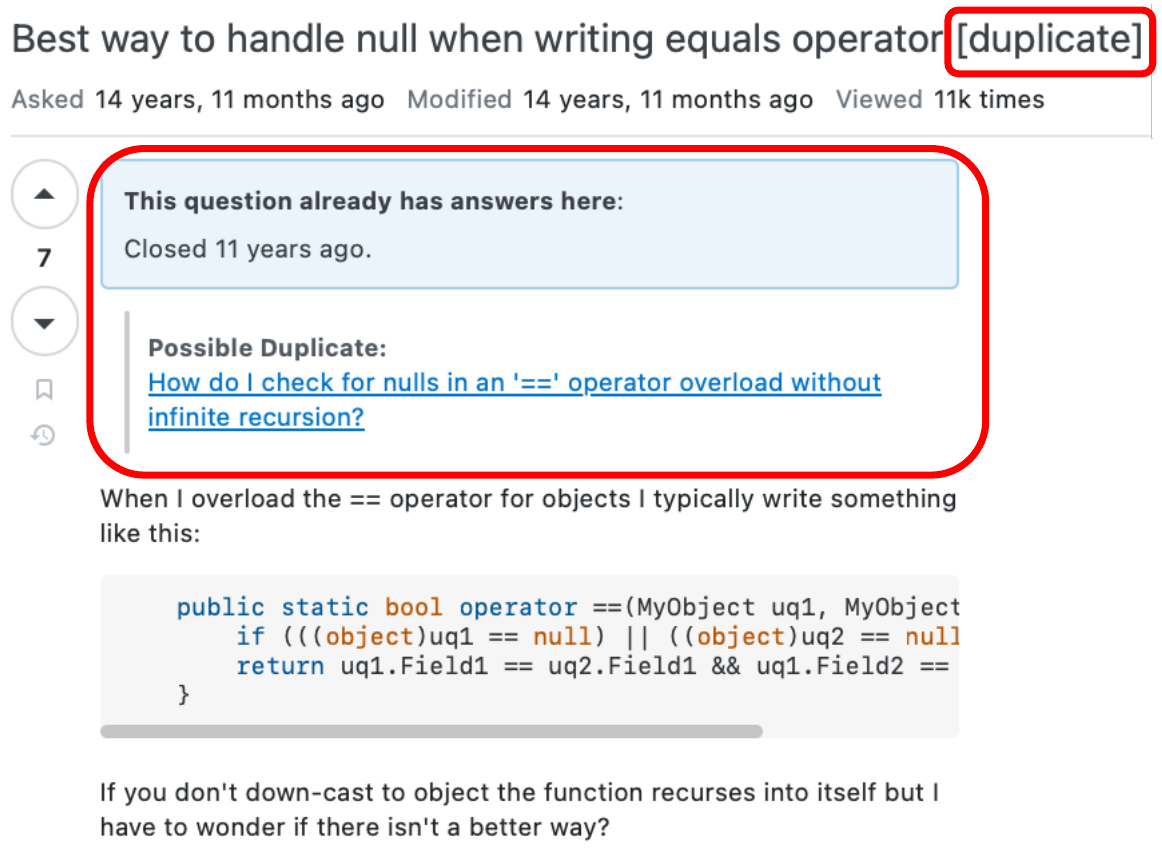}
    % \vspace{-8mm}
	\caption[Duplicate annotation]{An example of the duplicate annotation on Stack Overflow.}
 % question number: 86947
	\label{fig:dup_example}
    % \vspace{-2mm}
\end{figure}

\noindent\textbf{How do technical forums deal with duplicates?}
Online technical communities have their own mechanisms and regulations in place to handle duplication. For example, posts can be marked as duplicates in the question-closing procedure on Stack Overflow. A voting-based mechanism is adopted to mark the duplicate posts and link them to the original questions that already have at least an answer. Specifically, high-reputation users and moderators can vote to close a question as a duplicate. % and refer the duplicate to the original counterparts (one question may be linked to multiple previously asked questions)
\accepted{\heng{singular?}\xingfang{may not have only one duplicate. one questions may be linked to many previously asked questions.}\heng{added the explanation}} When a threshold of votes\accepted{\heng{a threshold of what}} is reached, the duplicate post will be closed and linked to the original ones with answers (one question may be linked to multiple previously asked questions), as shown in Figure~\ref{fig:dup_example}.\accepted{\heng{Refer to Figure 2 as an example, or move the figure in this section}} As the existing manual process involves much time and effort from the community members, timely identification of duplicate questions can be challenging, potentially causing delays for users in finding the correct answers~\cite{zhang2015multi, ahasanuzzaman2016mining}. \accepted{\heng{would be good to show some concrete numbers, e.g., an example post that takes a long time to be identified as a duplicate, or cite a source}} Moreover, many duplicate questions have been closed without a proper tagging of their duplicate nature\accepted{\Foutse{it is unclear what you are trying to communicate here! do you aim to say that they are being closed without a proper tagging of their duplicate nature? You might want to clarify this!}}, which causes redundancy and increases the users’ burden of filtering information~\cite{zhang2015multi}.\accepted{\heng{would be good to provide an example or cite a source}} By referring to the answers of these duplicate questions that have not been identified and assoicated yet, users are more likely to find solutions to their problems quickly.
\accepted{\Foutse{reference? or elaborate...what do you mean by aggregation here exactly? labeling duplicates?}\xingfang{I changed the wording. I used the word 'aggregation' as we planed to do summarization over answers.}}

\noindent\textbf{The need for automated detection of duplicate posts.}
The voting-based approach adopted by Stack Overflow demands heavy human efforts and may not work efficiently, given the number of new posts emerging every day. Automatically detecting duplicates in technical forums can help improve working efficiency and benefit users by shortening the waiting time for solutions when their questions are not new (i.e., by linking a new post to an existing post with solutions). Moreover, there are a large number of duplicate questions which are not identified on Stack Overflow~\cite{zhang2015multi, ahasanuzzaman2016mining, kamienski2023analyzing}. Identifying these duplicates may reduce the forum’s redundancy and improve the online forum’s efficiency. 
Therefore, our study proposes an approach for automated detection of duplicate posts that is accurate, scalable, and requires minimal effort to pre-process or feature-engineer the post data.

\section{Related Works}  \label{sec:related_work}
Duplication has been an enduring and persistent issue in dealing with software-related textual data (e.g., bug reports, technical posts). In this section, we briefly introduce previous works that are closely related to our study, which includes text embedding, duplicate bug report detection and duplicate post detection.

% \xingfang{Washi sensei: general: different author. Near-duplicate document detection. documentation by different authors. why we have to address?}

\subsection{Text Embeddings and Semantic Similarity}
Bengio et al.~\cite{bengio2000neural} first proposed neural embeddings in the form of a feed-forward neural network language model. Neural embeddings are dense vector representations of texts learnt by different neural network structures and objective functions. Word embedding techniques (e.g., Word2Vec~\cite{mikolov2013distributed}, GloVe~\cite{pennington2014glove}) can produce high-quality vectors that can represent the semantic meaning of words, which are then extended to word sequences (i.e., sentence embeddings or document embeddings~\cite{le2014distributed}).
These embedding techniques have been applied to various NLP applications (e.g., machine translation~\cite{li2014neural, zhang2014bilingually}). LLMs (e.g., BERT~\cite{devlin2018bert}, GPT-3~\cite{brown2020language}), which are trained on massive corpora with complex neural network structures, have attracted substantial attention in recent years. They can capture contextual information and operate on different levels of granularity. They provide a new approach for solving different natural language downstream tasks and have achieved superior results. Traditional approaches to measuring the similarity between two texts are usually based on linguistic and statistical features~\cite{wang2020measurement}. Nonetheless, the advent of text embeddings now enables us to measure the similarity between a pair of texts with the embeddings and distance functions, which enables applications on software-related textual data~\cite{xu2021post2vec}.
In this work, we use the GPT-3 model to generate initial semantic embeddings for technical posts.\accepted{\heng{relate them to our work}}

\subsection{Detecting duplicates of bug reports} 
Bug reports generated from various testing and development activities are prone to duplication. Duplicating bug reports leads to increased maintenance efforts during bug triage and fixing throughout the software lifecycle~\cite{kucuk2021characterizing}. Multiple approaches have been developed to detect bug report duplicates. Similar to technical forum posts, bug reports are structured natural language, which comes with a certain format or structure and may include elements like code blocks.\accepted{\heng{need to explain what is ``structured natural language''; besides, bug reports, like forum posts, can include other elements such as code}} Therefore, existing approaches are mainly based on performing feature engineering on textual data and assessing the similarity between pairs of bug reports~\cite{sun2011towards,zhang2023duplicate}\accepted{\heng{seems we need more than one citations here}}. For example, Sun et al. proposed SEP~\cite{sun2011towards}, which is a retrieval function to rank bug reports by their similarities\accepted{\heng{rank by what}}. This approach takes into account both textual and categorical features of bug reports. It then leverages a weighted linear fusion of these factors to compute the similarity between pairs of bug reports. Textual similarity is computed with a variant of the BM25 ranking function~\cite{robertson2004simple}.

Besides, deep learning approaches are also employed by researchers to encode the features of bug reports. Siamese Pair~\cite{deshmukh2017towards} is a deep learning-based architecture where different fields of bug reports are encoded by different sub-structures.
% (e.g., a summary of a bug report is encoded with a Bi-LSTM structure, bug description is encoded with a CNN)\heng{cite}\heng{this is not the concept of the siamese structure, but multi-modal structure? A siamese structure has two identical sub-neural networks so that it can classify positive and negative pairs?}. 
Finally, the encoded features are concatenated and a Siamese structure is used to train and refine the encodings. Moreover, the semantic embeddings are also introduced in the bug report duplicate detection task. He et al.~\cite{he2020duplicate} proposed DC-CNN, which employs a word2vec model to capture the semantic information in different fields of bug reports. Inspired by these duplicate detection approaches on bug reports, we adopt a Siamese-based network structure to refine the embeddings generated by an off-the-shelf pre-trained language model.
\accepted{\heng{Explain how our work is related to them (e.g., inspired by XX study, we use a siamese structure to ...).}}

\subsection{Detecting duplicates of technical posts}
In the field of mining technical forum data, while fewer studies are dedicated to duplicate detection, several strategies have been investigated to address the challenge of identifying duplicates. As the tasks exhibit resemblances, the current methods exhibit certain similarities with those employed for bug reports. Zhang et al.~\cite{zhang2015multi} proposed DupPredictor, which is an automated duplicate detection approach that considers multiple factors similar to the above-mentioned SEP~\cite{sun2011towards}\accepted{\heng{cite}}. This approach calculates the similarities between two technical posts using the titles, descriptions, and tags of the posts. A post pair's topic distribution is captured using a topic modeling algorithm, and in conjunction with various similarity metrics, a composer is utilized to combine these factors, which results in a final score that determines the degree of duplication. In contrast, Dupe~\cite{ahasanuzzaman2016mining} uses five carefully selected features, which include cosine similarity, term overlap, entity overlap, entity type overlap, and WordNet similarity, to construct a discriminative binary classifier to detect duplicates for technical posts. %After training, the classifier exhibits better performance than DupPredictor. 
Wang et al.~\cite{wang2019detecting} explore deep learning approaches based on three structures (i.e., CNN, RNN, and LSTM) to solve the problem of duplicate post-detection on Stack Overflow. %The experimental outcomes demonstrate the superior performance of the LSTM-based model in comparison to traditional methods such as DupPredictor and Dupe. 
Subsequently, in a later study, they enhanced the performance by incorporating Word2Vec embeddings~\cite{wang2020duplicate}. In addition, some works (e.g., Kamienski et al.~\cite{kamienski2023analyzing})\accepted{\heng{if mentioned multiple works then cite multiple, other wise use e.g.}} focus on some specific domains and evaluate different approaches on some domain-specific forum post datasets. Although these approaches claim good performances on some datasets, they still suffer from the lack of generalizability and scalability, which refrains them from practical uses. Our endeavor is dedicated to remedying these limitations while updating the research domain with assessments based on more recent technical forum data.
\accepted{\heng{Explain the limitations of existing works (why our work is needed at all) and clarify the novelty of our work over the existing ones (why we can address the limitations).}}

\section{Dataset Construction} \label{sec:data_construction}

Most of the existing duplicate post datasets are of limited size and outdated. Therefore, to enable better evaluation, we constructed our duplicate technical post dataset from a recent Stack Overflow dump containing all questions existing on December 6th, 2022\footnote{https://archive.org/details/stackexchange}\accepted{\heng{cite the data dump}\xingfang{https://archive.org/details/stackexchange}}. We used SQL query to access IDs of all the duplicate posts, and with the IDs of all related posts, we extracted the title, main body and other meta information from the data dump. 

\begin{figure}[!ht]
    \centering
    % \vspace{-3mm}
	\includegraphics[scale=0.4]{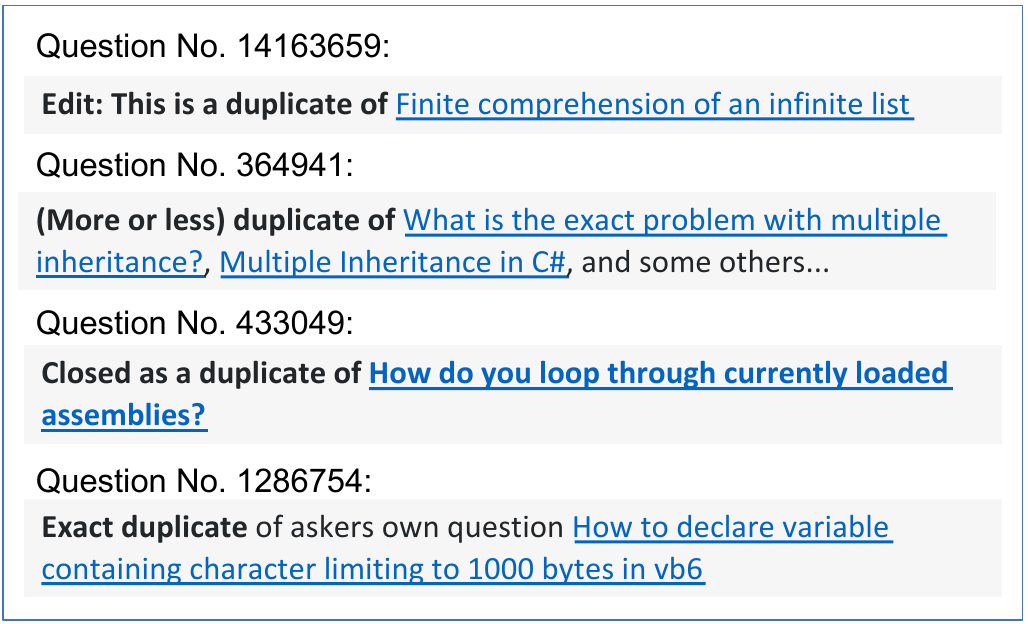}
    % \vspace{-3mm}
	\caption[Duplicate annotation]{The examples of customized annotations on Stack Overflow.}
	\label{fig:dup_cus_ann}
    % \vspace{-3mm}
\end{figure}

In the title and body of some duplicate posts, some special tokens and blocks are used to mark and link them to the original posts. For example, a typical duplicate annotation on Stack Overflow is shown in Figure~\ref{fig:dup_example}. When a post is closed as a duplicate, the title will be appended with the text \textbf{“[duplicate]”} and a block will be added in the body of the post to show the possible duplicates\accepted{\heng{duplicates?}}. However, some posts may be closed and edited by users or moderators with customized annotations appended to the end of the posts\accepted{\heng{not sure what ``end'' means here}}. Figure~\ref{fig:dup_cus_ann} shows some examples of these customized annotations.

The forms and formats of duplicate annotation may vary across posts. In particular, these tokens and blocks may leak the annotation to the duplication detection algorithms and lead to unreliable evaluation results. Therefore, we remove these tokens and blocks with regular expressions designed for the different formats.
% \heng{This is a bit confusing here: such annotations are used in the majority of the study, only removed when deriving the features for training and testing the models, right? It would be less confusing to only mention it later in the implementation details.}\xingfang{it is about constructing the dataset to ensure no label leaking to the data.}
%\xingfang{We didn't use these annotation because including these annotations would cause data leaking.} \heng{I meant removing the tokens are just for training/evaluating the model, but we use the information for other purposes like the preliminary study (thus we still have the information). Maybe be more specific and say we remove these the tokens in the dataset used for training and evaluating the approach?}\xingfang{No, we didn't use these blocks. There is a special table in their database to mark the duplicate relations.}
% need to show an example
%\heng{Some important information is missing: 1) how duplicate posts are extracted (you explained how SO marked them, but not how you extracted the information) (e.g., use regular expression); 2) which domains (tags?) are selected and why.}
%\xingfang{1) the ids of duplicate posts are in their database. we just use a sql query to acquire them, as described in the first paragraph of this section. 2) we extracted posts of all domains.}
With the data dump, we were able to extract 723,008 pairs of duplicate posts. It is worth noting that one post may be a duplicate of more than one post and has more than one duplicate counterpart.
% \heng{we need a clear definition of what is a duplicate to avoid confusion}.\accepted{\heng{mentioning this is too early (should be mentioned when/after explaining model building and evaluation)}}

% \secdescription{Time/No.Q/No.Dup/Labels}
% \secdescription{Pre-processing: remove blocks/ remove duplicate annotations -> in case of data leaking}

% \subsection{Qualitative Study on Duplicate Posts on Stack Overflow}
\section{Preliminary Study} \label{sec:preliminary}

% \secdescription{tags with the largest amount of duplicate, also percentage of duplication}

% \secdescription{Maybe compare with previous census to see the variation over time. More duplicate?}

% \secdescription{Findings: duplicate questions usually have common labels. only 4.49\% do not have even one common label. -> common label along with timestamp can serve as a filter in the process of duplicate detection.}

To comprehend the current status of the marked duplicate posts on Stack Overflow, we briefly census
% undertake a brief qualitative \heng{it's a quantitative analysis?} study using
our extracted dataset. This census aims to examine the distribution of duplicates based on their tags. These tags are specific keywords or labels that relate to a particular field or technology being discussed. A post is categorized under a technical tag if its tag list includes that particular tag. Therefore, a post may belong to multiple categories, and these categories are not mutually exclusive in our study. As users are allowed to input their customized tags, the total number of distinct tags is significant, amounting to 29,035 in our dataset\accepted{\heng{in our studied data dump?}}. We refer to these tags as technical \textbf{topics} to emphasize that they represent specific subject areas in our research. 

We list the Top-20 topics with the largest numbers of posts in Table~\ref{tab:top20}.\accepted{\heng{``domain'' is used in the table while ``tag'' is used in the text: link needs to be built}} The results show that JavaScript-related posts rank first in the number of duplicate pairs, with 122,968. However, string-related posts have the highest rate of duplication: among all the questions with the ’string’ tag, around 20\% of posts are duplicates and their original questions. We compare the most popular tags with those of the previous studies~\cite{ahasanuzzaman2016mining}. The list of the Top-20 popular tags remains largely in line with the previous findings, although there are slight differences among the tags that are most commonly used. The tags 'string,' 'array,' and 'regex' continue to hold prominent positions in the ranking, suggesting that some fundamental programming questions are repetitively asked by programmers. Moreover, 'pandas' and 'r,' which were not present in the previous ranking of popular tags, have emerged in the latest results. This could indicate a growing popularity on and interest in the field of data science. The variations in tag ranking indicate the shifts in the distribution of the Stack Overflow data changes over time, which underscore the necessity for keeping research data up to date.
\accepted{\heng{I feel the popularity of the tags is not important, may be briefly mentioned}\xingfang{modified the paragraph.}}

\begin{table}
\centering
\caption{Top-20 topics with the largest numbers of posts. \accepted{\heng{We can call it ``topic'' instead of ``domain'' throughout the paper (e.g., regex is not a domain). According to SO's definition: A tag is a word or phrase that describes the topic of the question.}} \accepted{\heng{``No. of pairs'' to ``Dup. pairs''; ``No. of posts'' to ``Dup. posts''}}}

% \vspace{-3mm}
\label{tab:top20}
\resizebox{\linewidth}{!}{
\begin{tabular}{l|cccc} 
\hline
Topic     & \begin{tabular}[c]{@{}c@{}}Dup. posts\end{tabular} & Dup. pairs   & Total posts & Ratio    \\ 
\hline
string     & 37,217                                                             & 31,808  & 179,991      & 20.68\%  \\
arrays     & 55,329                                                             & 43,603  & 403,843      & 13.70\%  \\
regex      & 34,135                                                             & 20,863  & 254,924      & 13.39\%  \\
c          & 48,570                                                             & 29,699  & 390,087      & 12.45\%  \\
c++        & 88,502                                                             & 50,517  & 783,567      & 11.29\%  \\
r          & 53,124                                                             & 26,890  & 473,420      & 11.22\%  \\
pandas     & 27,634                                                             & 15,412  & 267,949      & 10.31\%  \\
python     & 209,759                                                            & 113,278 & 2,071,327    & 10.13\%  \\
java       & 175,586                                                            & 100,149 & 1,878,116    & 9.35\%   \\
css        & 69,474                                                             & 39,041  & 779,102      & 8.92\%   \\
javascript & 210,869                                                            & 122,968 & 2,453,713    & 8.59\%   \\
python-3.x & 28,185                                                             & 25,682  & 329,120      & 8.56\%   \\
php        & 118,937                                                            & 66,656  & 1,451,348    & 8.19\%   \\
html       & 85,293                                                             & 62,859  & 1,156,016    & 7.38\%   \\
mysql      & 43,750                                                             & 28,409  & 655,003      & 6.68\%   \\
c\#        & 103,481                                                            & 59,106  & 1,571,198    & 6.59\%   \\
sql        & 37,167                                                             & 27,008  & 650,208      & 5.72\%   \\
jquery     & 50,569                                                             & 35,933  & 1,031,378    & 4.90\%   \\
android    & 57,728                                                             & 34,102  & 1,393,126    & 4.14\%   \\
ios        & 26,873                                                             & 18,087  & 677,311      & 3.97\%   \\ 
\hline
Sum        & 1,562,182                                                          & 952,070 & 18,850,747   & 8.29\%   \\
\hline

\end{tabular}
}
% \vspace{-5mm}
\end{table}

We also do a census over the duplicate pairs. We find that \textbf{there is at least one common tag in 95.5\% of the duplicate pairs}, which means that common tags may be a good indicator for duplicates. This finding justifies the practices of using tags as a feature to detect duplicates or measure the similarity of posts by previous works~\cite{zhang2015multi, ahasanuzzaman2016mining, zhang2017feature}. Within the posts that are annotated to have duplicates, on average, each post has approximately 1.07 duplicate counterparts. Notably, \textbf{the majority of these posts, around 93.7\%, have just one duplicate}, while a smaller percentage, roughly 6.3\%, have more than one duplicate. We consider duplication a transitive relation. Based on the existing annotations, we found that at least 5\% of the labelled duplicate posts are not linked with all their duplicate counterparts. This highlights an issue of incomplete duplicate annotations on Stack Overflow.
% we found that around 5.0\% of the labelled duplicate questions are not really duplicates of their counterparts. Therefore, the quality of duplicate annotations presents an issue on Stack Overflow.

\accepted{\heng{Would be interesting to discuss the time gap between the duplicate pairs, how soon a post is marked as a duplicate after its creation, and how many the number of duplicates per each post. Such information can help motivate our work: e.g., if it takes a long time to mark a duplicate after its creation, it means manual annotation is not efficient; if a post is a duplicate of multiple posts, it means duplication is a serious problem. }\xingfang{modified.}}

\section{Detecting Duplicate Questions\accepted{\heng{no need to highlight GPT-3 in title}}}
\label{sec:method}

\begin{figure*}[!htbp]
    \centering
    % \vspace{-3mm}
	\includegraphics[scale=0.60]{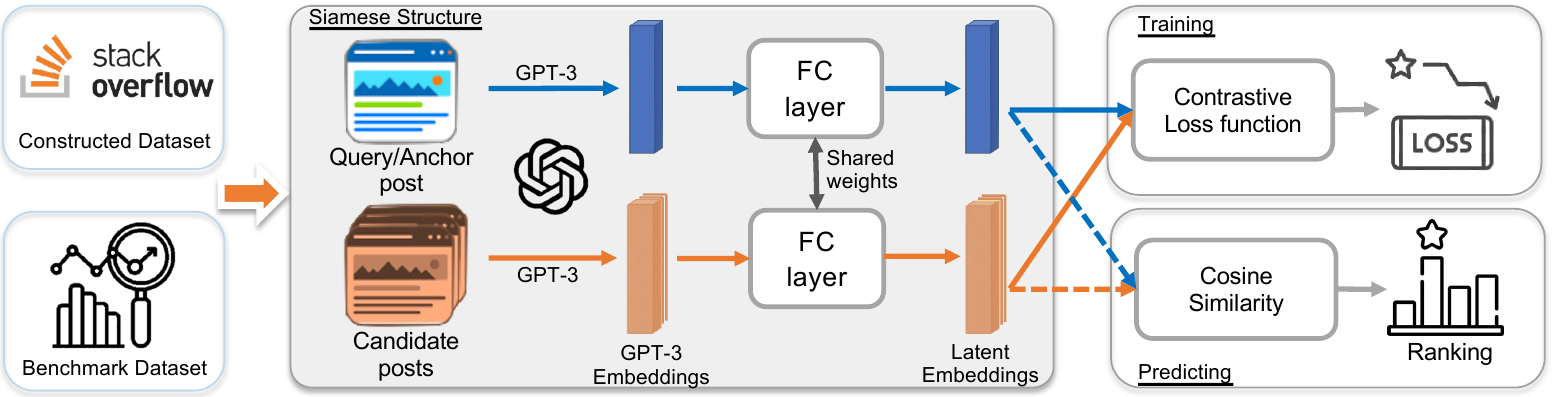}
    % \vspace{-3mm}
	\caption[Overview]{An Overview of our proposed approach.}
	\label{fig:overview}
    % \vspace{-3mm}
\end{figure*}

\subsection{Overview}

\accepted{\heng{Bring Figure 1 here to guide the description?}}
In this section, we present a detailed account of our proposed approach for detecting duplicate technical posts on Stack Overflow. Firstly, we provide the details of our embedding generation process. Then, we elaborate on the Siamese-based network structure for refining the GPT-3 embeddings. Finally, we explain the two loss functions we adopt to train the Siamese-based network structure to obtain the refined latent embeddings. These components collectively contribute to our strategy for refining the GPT-3 embeddings using the available annotations for detecting duplicate posts.

\subsection{Embedding generation} \label{sec:emb_gen}
\accepted{\heng{Explain the unit of data used to generate the embedding (each post, right?), the information used (title, body, answer?). Explain why not using code? The input dimension limitation of GPT-3 is not a good reason: we don't want to use tool to limit our problem, but to choose the tool based on our problem. It's better to cite existing work to support the choice of not using code (e.g., using code may not improve the performance, or prior work does not use code).}\xingfang{just simply title and body. no answer. not using code block is one of the threat to validity.}}
In the dataset construction stage, we preprocess the post data by removing the HTML tags and unnecessary blocks from the raw Stack Overflow data. We use the \textit{text-embedding-ada-002} model, which is a GPT-3 embeddings model from OpenAI and exhibits superior performance over previous versions in our experiments. We concatenate the title and description of posts and acquire the embeddings using the API provided by OpenAI. The generated embeddings have 1536 dimensions. The max input tokens limit of this model is 8191 tokens. However, some users of Stack Overflow copied unwieldy artifacts (e.g., lengthy urls with base64 encoding, codes or other program-generated texts) to the body part of posts, leading to many lengthy posts, which exceed the length limit of the model input. The inclusion of these bulky data tends to hinder the embedding process due to their length and limited contextual relevance. Therefore, we remove these unexpected texts using some regular expressions designed for these cases. This helps maintain the focus on the pertinent information and ensures better embedding quality. In some very rare cases, where concatenated texts are still too long to be the model input, we just truncated the input to the allowed maximum length. In the embedding generation process, we prepare the input for the GPT-3 model by concatenating the title and body of posts. We send the concatenated texts to the GPT-3 model using the API provided by OpenAI. 

\subsection{Siamese structure for refining the GPT-3 embeddings}
The GPT-3 embeddings possess the capability to encapsulate semantic meaning within the technical forum posts. Therefore, we can use the off-the-shelf embeddings generated with the GPT-3 model to evaluate the semantic similarities between posts, which may stand a good chance of finding duplicate relations among technical posts. Under this unsupervised manner, we can solve the duplicate detection task. Nonetheless, there exists a potential to enhance these embeddings by leveraging annotations readily accessible from the Stack Overflow data dump. Although these annotations may not be exhaustive, as some duplicates remain unmarked, leading to false negatives, the positive annotations are reliable due to their validation through manual checks by multiple users. In this study, we introduce a Siamese network~\cite{bromley1993signature} to further refine the GPT-3 embedding for representing duplicate relationships in posts by utilizing the positive pairs of duplicate posts.

The Siamese-based structure is illustrated in the gray background portion of Figure~\ref{fig:overview}. The inputs to the network are the embeddings of posts. The two network structures share the same parameters, and therefore, the structure is called Siamese. Based on the outputs generated by the network structures, we calculate the loss and update the models' parameters during the training process. When predicting, We compute the cosine similarity between posts using the latent embeddings generated by the outputs of the network structures. We then sort the candidate posts in descending order of similarity score. In our experiments, we adopted a fully-connected layer with 50 nodes as the Siamese structures. The choice of the structure serves as a hyperparameter of our approach.

\accepted{\heng{more details needed to explain how the latent embeddings are trained}}

% \begin{figure}[!h]
%     \centering
% 	\includegraphics[scale=0.45]{./siamese_structure.pdf}
% 	\caption[Siamese]{The Siamese structure. \xingfang{Temporary one. Will update later.}\heng{The figure should explicit show ``Positive pairs'' and ``Negative pairs''. Check existing figures for siamese structures.}}
% 	\label{fig:siamese}
% \end{figure}

\subsection{Loss functions}
\label{sec:loss}
We aim to optimize the proposed Siamese-based network structure to strengthen the distinction between the latent embeddings of\accepted{\heng{duplicate pairs and non-duplicate pairs?}} duplicate and non-duplicate posts, making them more distant while bringing the embeddings of duplicate posts closer together. To achieve this objective, we utilize the following two loss functions, which are designed to promote the intended behavior. It is worth noting that we use the cosine distance as our distance function and optimization objectives for loss functions as recommended by OpenAI\footnote{https://platform.openai.com/docs/guides/embeddings/limitations-risks}.

%\heng{I remember you considered another simpler loss function before moving to triplet loss? we can show the results for that simpler loss before moving to the triplet loss, to show the benefits of using the triplet loss.}
%\xingfang{the cross entropy (binary classification) loss hardly work, not proper to mention.}

\textbf{Triplet Loss}~\cite{hoffer2015deep} is widely applied to tasks like image retrieval, face recognition, and similarity learning where image and text embeddings need to be optimized. For each training sample, this loss function demands three data points\accepted{[explicitly explain what data points mean in our context]}: an anchor, a positive and a negative sample\accepted{\heng{add a noun here}}. Each sample is a post in our case. The central idea behind this loss function is to simultaneously minimize the distance between an anchor and a positive sample and maximize that of the anchor and a negative sample. The anchor serve as a reference for the model to differentiate between similar and dissimilar posts. \accepted{\heng{explain why we want to obtain this (what an anchor means here)}} Mathematically, the triplet loss can be defined as Equation~1.

% \vspace{-3mm}
% \begin{equation*}
% \label{eq:triloss}
% % \vspace{-1mm}
% \begin{aligned}
%     \text{Triplet Loss} =
%     \max(0, \ &||f(\text{Anc}) - f(\text{Pos})||^2\\
%     &- ||f(\text{Anc}) - f(\text{Neg})||^2 + \text{margin}) \  (1)
% \end{aligned}
% \end{equation*}
\begin{equation*}
\label{eq:triloss}
\begin{aligned}
    \text{Triplet Loss} = \max(0, & \ \ dist(f(\text{Anc}), f(\text{Pos})) \\
    & - dist(f(\text{Anc}), f(\text{Neg})) + \text{margin}) \ (1)
\end{aligned}
\end{equation*}
\accepted{\heng{always provide an equation id to refer to (for us and for reviewers/readers)}}

\noindent where $f(x)$ denotes the embedding of a technical post and $dist(x)$ denotes the distance formula, which is cosine distance in our case. The margin is used to enforce that the model correctly identifies the anchor and positive posts as similar and ensures that they are sufficiently dissimilar from the negative posts by a minimum acceptable separation.\accepted{\heng{keeping the pairs apart from each other, or the posts within each pair apart from each other. Please use explicit concepts in this context}}

In our constructed dataset, there are only duplicate annotations for candidates for a certain query post. Therefore, we assume that posts not marked as duplicates of the query post are not duplicates of the query, which ignores unmarked duplicates in the dataset. When constructing a triplet training sample, we randomly select a post from these unmarked posts from the dataset as a negative sample.

% In our constructed dataset, there are only duplicate annotations for candidates for a certain query post. Therefore, we assume that posts that are not marked as a duplicate for the query postare not duplicate of the query, which ignores unmarked duplicates in the dataset. We randomly select a post from these unmarked posts\accepted{\heng{without a duplicate mark for the target post?}} from the dataset as negative sample when constructing a triplet training sample.

\textbf{Multiple Negative Ranking (MNR) Loss}~\cite{henderson2017efficient} is an extension of the Triplet loss. It differs from the Triplet Loss in that it considers multiple negative examples for each anchor. This loss operates on a batch of post pairs 
$(a_i, p_j)$, where ($0\leq i,j<N $) and $N$ is the batch size. When $i = j$, the post pair is a duplicate pair, which means $a_i$ is duplicate of the post $p_j$. While $i \neq j$, the pair is a negative pair in a batch. For each $a_i$, the loss function uses the post $p_j$ ($i = j$) as positive sample and all $p_j$ ($i \neq j$) as negative samples and minimizes the negative log-likelihood for softmax normalized scores. As we adopt the random batch selection, $N-1$ randomly selected negative samples are used in the learning process. Therefore, the batch size $N$ serves as a hyperparameter for both the training setting and the loss function.

% $(a_1, p_1)$, $(a_2, p_2)$, ..., $(a_N, p_N)$, where $N$ is the batch size and $(a_i, p_i)$ where  represent positive pairs (i.e., $a_i$ and $p_i$ are a pair of duplicate posts) and $(a_i, p_j)$ ($i \neq j$) signifies negative pairs (there are $n-1$ negative pairs in a batch). It samples in each batch n-1 negative instances randomly\heng{this sentence probably needs rephrase with some context}. For each $a_i$, it uses all $p_j$ as negative samples and minimizes the negative log-likelihood for softmax normalized scores.\heng{not clear how the loss function is calculated when there are multiple negative functions. maybe use an equation?} Therefore, the batch size serves as a hyperparameter for both the training setting and the loss function.
% \heng{it's a bit confusing, why (a1, p1), (a2, p2)..., seems (a1, p2) or (a2, p1) etc. do not exist? the index needs some explanation here}

\section{Evaluation} \label{sec:eval}

%\subsection{Overview}
% In this section, we aim to assess the performance of our approach, and evaluate the learned latent embeddings's capability of representing duplicates of technical posts and understand its characteristics and effectiveness in detecting unlabelled duplicates. We achieve our research objectives by answering the following Research Questions (RQs).

% \heng{it's not clear it's about the GPT-3 embedding or the embedding resulting from our approach. I would suggest using another term to distinguish our contribution from GPT-3}

In this section, our goal is to assess the performance of our approach and examine the latent embeddings' capability to represent duplicates of technical posts. We also aim to gain an understanding of their characteristics and evaluate their effectiveness in detecting unlabeled duplicates, all of which will be achieved by addressing the three Research Questions (RQs) in this section.

\subsection{Evaluation datasets}

In our study, we utilize two datasets to assess our approach: a publicly available benchmark dataset, CQADupStack~\cite{hoogeveen2015cqadupstack}, and a dataset we constructed as detailed in Section~\ref{sec:data_construction}. The use of the former dataset allows us to benchmark our performance against prior studies, whereas employing our constructed dataset, which is both larger and more up-to-date, provides us with a more comprehensive understanding of the performance and characteristics of our approach.
\accepted{\heng{Briefly describe the main evaluation dataset, with reference to the Dataset Construction section.}}

%\subsection{Baselines and benchmark}

\subsection{Evaluation metrics} 
% \heng{Move to the Evaluation section}
% As we formulate duplicate detection as a ranking task, we employ the Top-N accuracy to evaluate embeddings’ ability to detect duplicates.

\noindent \textbf{Top-N accuracy} selects the topmost k candidates from the ranking results based on the distances of their embeddings compared to those of the input queries. It calculates the frequency with which the correct labels appear among the predicted top k labels. We used Top-N accuracy to evaluate the duplicate detection approaches under a ranking formulation.

% \noindent \textbf{Top-N accuracy} selects the topmost k candidates from the ranking results based on the distances of their embeddings compared to those of the input queries. It calculates the frequency with which the correct labels appear among the predicted top k labels.

% It computes the number of times the correct labels\heng{or ``at least one correct label (i.e., duplicate posts)''} are among the predicted top k labels.

% \heng{Also consider MRR and MAP?}

\noindent \textbf{Area Under Curve (AUC)} assesses the extent to which models can distinguish between different categories. When the duplicate detection is formulated as a classification task, the AUC indicates the probability that an approach can give a higher score (similarity or classification probability) for a duplicate post pair than a non-duplicate pair.

% assesses the extent to which models can distinguish between different categories. While we initially frame duplicate detection as a ranking problem, for the sake of conducting more comparisons with other methodologies, we convert our approach into a classification task. In this classification scenario, we evaluate the models' capability to discern duplicates by measuring the similarity between two input posts. To evaluate these models, we employ the Area Under Curve (AUC) score, which gauges their proficiency in discriminating between duplicate and non-duplicate pairs.

% measures the degree of separability of models. Although we formulate the duplicate ddetection as a ranking problem, to ensure better comparisons with other approaches, we transform our approach into a classification one, in which we calculate the similarity between two input posts and evaluate its ability to recognize duplicates. We adopt the Area Under Curve (AUC) score to evaluate models’ abilities to distinguish duplicate and non-duplicate pairs.

% As our baseline essentially adopts the classification formulation, in order to compare our approach with the baselines\accepted{\heng{one or multiple baselines?}},

\accepted{\heng{our approach is more than the embeddings, don't limit to the embeddings as only using the embeddings do not indicate a significant contribution}}

\subsection{Comparing with baselines} %\heng{Shall we move the comparison with baselines to RQ1, as we only use the baselines in that RQ?}\xingfang{I think it is fine to leave it here. Or the structure (length of each section) will be very unbalanced. And it is fine to serve as an introduction for the RQs, even though the latter two RQs do not use it.}

% \xingfang{TODO: indicate we compared with the best supervised results across all the variants in the benchmark dataset paper. Compared with duppredictor over ranking? Each topic is trained separately.}

\noindent\textbf{Benchmark dataset used for comparison.}
To enable better comparisons with previous duplicate detection approaches, we conduct comparative experiments on CQADupStack,\accepted{\heng{update: we only use the dataset for comparison, not our main dataset}} which is a benchmark dataset for community question-answering (cQA) research~\cite{hoogeveen2015cqadupstack}. This dataset comprises fixed training and testing data, and some recent approaches~\cite{lau2016empirical, zhang2017detecting} reports their performances on this benchmark. Therefore, we employ CQADupStack as the benchmark dataset. The CQADupStack dataset is collected from the Stack Exchange dump over twelve sub-forums. We followed the practice of a previous work~\cite{zhang2017detecting} to use posts from sub-forums over nine topics that are closely related to software development and programming language to conduct our experiments to enable fair comparisons. We use the training set to search for the optimal parameters for the DupPredictor, and train the Siamese network in our approach. All approaches evaluated by us are evaluated with the test set that contains 17,149 pairs, among which 10\% are duplicate pairs. Despite the benchmark dataset's limited size and its construction dating back some time, its use allows us to conduct a comparative analysis of our approach against the baseline methods.

% We further evaluate our approach with our constructed dataset, which is more up-to-date and of a larger size.

% As there is no public available implement for our

\accepted{\heng{It reads like our evaluation is only performed over the benchmark. Is it true? If this benchmark dataset is only used for comparing with the baselines, then we probably need to clarify it}.
\heng{I suggest to name this sub-section as ``Baselines and benchmark'', in which we first introduce the baselines, then explain how the comparison is done using the benchmark.}}

\noindent\textbf{Baseline approaches.}
Prior researchers~\cite{zhang2015multi, ahasanuzzaman2016mining} evaluate their duplicate detection approaches with different datasets, which are constructed with publicly accessible Stack Overflow data dumps at the time of their studies.\accepted{\heng{available at the time of performing the experiments?}}\accepted{\heng{Would be good to give an example}\xingfang{the cited papers.}} However, the continuous updates to the Stack Overflow data dump pose challenges for reproducing experimental outcomes and conducting unbiased comparisons based on the reported results, given that most original datasets are no longer available.\accepted{\heng{Maybe don't say it's challenging for reproducing the results (one can just use the exact old data dump; besides, we also compare our results with one baseline reported in the original paper). Instead, we can say that the results may not be reproducible on the new (evolved) data}} Moreover, their results may not be reproducible on new data: A reproducibility study~\cite{silva2018duplicate} mentions that the barriers to replicating previous approaches are significant. During the reproduction, the authors found that certain methods may exhibit lower performance when different datasets and implementations are utilized.

We compare our proposed method with three baseline techniques, DupPredictor~\cite{zhang2015multi}, Doc2vec (dbow)~\cite{lau2016empirical} and PCAQDup~\cite{zhang2017detecting}.
% \accepted{\heng{Can we say these two baselines are the state-of-the-art approaches or the best-performing approaches?}\xingfang{hard to say, all paper claim that they are SOTA. these two are not newest ones.}\heng{we need a rationale for choosing these two baselines but not others...}} \accepted{\heng{represent two types of existing techniques for duplicate post detection?}}\accepted{\heng{the difference is not clear, as embeddings can be used for similarity measurement as well. Do you mean similarity measurement based on non-semantic features?}}
These three methods represent three types of existing techniques for duplicate post detection: DupPredictor mainly relies on similarity measurement of elaborated textual features. Doc2vec is a text embedding technique, while PCQADup combines the text embedding with textual features. 
We selected these baseline techniques because they either have publicly shared implementations for our reference or have published results on the benchmark dataset. We re-implement DupPredictor since the original paper did not provide their implementation. We re-implemented DupPredictor based on the replication package provided by Silva et al.~\cite{silva2018duplicate}.\accepted{\heng{We re-implemented DupPredictor based on the replication package provided by ...? Otherwise it conflicts with ``they either have existing implementations''}} To assess the performance of Doc2vec and PCQADup, we rely on the previously reported scores~\cite{lau2016empirical, zhang2017detecting} on the benchmark dataset.

\accepted{\heng{be precise: we compare our method with the baselines on the dataset used by Lau et al.~\cite{lau2016empirical}}}

% \heng{Move the evaluation metrics from Section 6 to  this section}\xingfang{Done}

% \xingfang{TODO: dataset composition}

\subsection{Research questions (RQs)}
% \heng{as a full section, not a sub-section?}
% \xingfang{then the evaluation section will have very few content. especially if we move the 'comparing with baselines to RQ1'. Do we move it here? Or we change the 'comparing with baselines' to 'Baseline approaches'. and move the description of the benchmark dataset to RQ1?}\heng{Maybe change the section title "Evaluation" to "Evaluation Setup"? It's ok to keep Comparing with baselines in the Evaluation setup section. Another alternative is to remove the Evlauation section and move the content to the corresponding RQs.}
% \textbf{RQ1: How is the performance of pre-trained GPT-3 embedding on duplicate detection of technical posts?}

\noindent \textbf{RQ1: How does the performance of our proposed approach compare to that of baselines on the benchmark dataset?}
%\heng{How well can our approach detect duplicate posts?}

% \textbf{Objective:}
\paragraph{\textbf{Objective}} This research question aims to assess the effectiveness of our proposed approach in duplicate post detection and compare it with the baseline approaches. % over the benchmark dataset. 
%Our objective is to 
The results can help us understand the capacity of the latent embeddings generated by our model in capturing the similarity of technical content and identifying the duplicates of technical posts.
% The performance can also reflect the embeddings' ability to capture semantic meaning and contextual nuances of technical posts.\heng{the last sentence is not an objective if this RQ?}

\paragraph{\textbf{Approach}} %\heng{We should also show the results of our approach on the main dataset (not to compare with the baseline, but to show the performance (how good our approach is), right?} 
In this research question, we use the CQADupStack~\cite{hoogeveen2015cqadupstack} as the benchmark dataset. We use the method described in Section~\ref{sec:emb_gen}\accepted{\heng{the section number does not match}} to generate the embeddings for the posts in the dataset, and train the Siamese network with annotated training data and the MNR loss function. We formulate the duplicate detection task both as both a classification and a ranking task to enable comparisons with baseline approaches. 
% \xingfang{duplicate with the evaluation metrics.}
% In the classification scheme, we use the normalized cosine distance between post pairs \accepted{\heng{post pairs?}} as the prediction scores and calculate and report the AUC score (i.e., the AUC score measures an approach's ability to distinguish duplicate post pairs from non-duplicate post pairs\heng{right?})~\cite{lau2016empirical}\accepted{\heng{does the baseline approaches use a similar way to calculate AUC? Then we could cite them}}. 
Same as what previous works~\cite{lau2016empirical, zhang2017detecting} did, we calculate the AUC scores for each sub-forum in the benchmark. We also list the Top-N accuracy for DupPredictor and our method over studied topics on the benchmark dataset.
\accepted{\heng{Mentioned AUC, need to mention top-N accuracy as well}}

% \secdescription{2. benchmark dataset With baseline, AUC score as formulated as a classification problem.}
\begin{table}
\centering
\caption{AUC Scores for Duplicate Detection Comparison}
% \vspace{-0mm}
\label{tab:results_bm}
% \refstepcounter{table}
\resizebox{\linewidth}{!}{
\begin{threeparttable}
%     \begin{tabular}{c|c|c|c} 
% \hline
% \multirow{2}{*}{Topic} & \multicolumn{3}{c}{ROC AUC}                                                                    \\ 
% \cline{2-4}
%                         & Ours           & DupPredictor & \begin{tabular}[c]{@{}c@{}}Doc2vec \\(dbow)\tnote{1}\end{tabular}  \\ 
% \hline
% android                 & \textbf{0.992} & 0.990        & 0.990                                                      \\
% gis                     & \textbf{0.993} & 0.985        & 0.940                                                      \\
% mathematica             & \textbf{0.966} & 0.940        & 0.960                                                      \\
% programmers             & \textbf{0.987} & 0.968        & 0.930                                                      \\
% stats                   & 0.981          & 0.973        & \textbf{1.000}                                             \\
% tex                     & \textbf{0.989} & 0.964        & 0.950                                                      \\
% unix                    & \textbf{0.985} & 0.979        & 0.980                                                      \\
% webmasters              & \textbf{0.972} & 0.940        & 0.930                                                      \\
% wordpress               & \textbf{0.992} & 0.957        & 0.980                                                      \\
% \hline
% \end{tabular}
    \begin{tabular}{c|c|c|c|c} 
\hline
\multirow{2}{*}{Topic} & \multicolumn{4}{c}{AUC}                                                                              \\ 
\cline{2-5}
                       & Ours           & DupPredictor & \begin{tabular}[c]{@{}c@{}}Doc2vec \\(dbow)$^1$\end{tabular} & PCQADup$^2$  \\ 
\hline
android                & \textbf{0.992} & 0.990        & 0.990                                                     & 0.910    \\
gis                    & \textbf{0.993} & 0.985        & 0.940                                                     & 0.925    \\
mathematica            & \textbf{0.966} & 0.940        & 0.960                                                     & 0.800    \\
programmers            & \textbf{0.987} & 0.968        & 0.930                                                     & 0.940    \\
stats                  & 0.981          & 0.973        & \textbf{1.000}                                            & 0.910    \\
tex                    & \textbf{0.989} & 0.964        & 0.950                                                     & 0.705    \\
unix                   & \textbf{0.985} & 0.979        & 0.980                                                     & 0.948    \\
webmasters             & \textbf{0.972} & 0.940        & 0.930                                                     & 0.860    \\
wordpress              & \textbf{0.992} & 0.957        & 0.980                                                     & 0.890    \\
\hline
\end{tabular}
    \begin{tablenotes}
        % \small
        % \footnotesize
        \scriptsize
        \item [1] The results are taken from Lau and Baldwin~\cite{lau2016empirical}. The results of best performing variants of Doc2vec (dbow) are presented here. 
        % We tried to re-implemented this approach, but failed to achieve the comparable results.\heng{We re-implemented this approach but achieved much worse results (i.e., AUC from ... to ...).}\xingfang{AUCs are around 0.6. showing the AUCs may make it more suspicious.}
        \item [2] The results are excerpted from Figure 7 in Zhang et al.~\cite{zhang2017detecting}, The results of best performing variants are presented here.
    \end{tablenotes}
\end{threeparttable}
}
% \vspace{-3mm}
\end{table}
\paragraph{\textbf{Results and Analysis}}
Table~\ref{tab:results_bm}\accepted{\heng{table number does not match}} presents the AUC scores of our proposed approach and three baseline methods across nine technical topics in the benchmark. 
% The DupPredictor results are calculated using our implementation, while the results of Doc2vec and PCAQDup are sourced from Lau et al.'s paper~\cite{lau2016empirical} and Zhang et al.'s paper~\cite{zhang2017detecting}.\xingfang{we mentioned this in the previous section. and we have annotations on the result table.}
Table~\ref{tab:results_topn} presents the performance of DupPredictor and our approach in terms of the Top-N accuracy. We did not report the Top-N accuracy of the Doc2vec (dbow) approach because the prior work did not report the Top-N accuracy and our re-implementation could not reproduce the promising results reported in prior work~\cite{lau2016empirical}.
% \heng{it would be good to report the top-N accuracy of the re-implemented Doc2vec if results available}
% \xingfang{Modified. Please check:}

\textbf{Our proposed approach performs better and more stable than the baselines in distinguishing duplicate post pairs from non-duplicate pairs across different topics.}
%From the results, we can see 
Table~\ref{tab:results_bm}\accepted{\heng{table number does not match}} shows that our approach achieves better AUC than the baseline approaches: %, although the numerical gaps are not huge:
Our proposed detection approach achieves an AUC score between 0.972 and 0.993,  higher than that of the baselines for eight out of the nine topics. The results indicate that the latent embeddings obtained with our approach exhibit high distinctiveness in identifying duplicate posts. 
The results also demonstrate that the performance of our approach remains relatively stable across the nine technical topics. In contrast, the baselines' performances are less uniform, with certain topics exhibiting competitive AUC scores (e.g., Doc2vec with \textit{stats}), while others notably lag behind (e.g., Doc2vec with \textit{programmers}, PCQADup with \textit{tex}).\accepted{\heng{move this to the AUC paragraph?}\xingfang{moved.}}
% \xingfang{The discrepancy between PCAQDup and others is large. we may not be able to explain like the following.} \heng{it's ok, we just need to say that the difference between ours and the other two baselines is small and explain it}\xingfang{Okay}
However, the absolute differences between the AUC scores achieved by our approach and those achieved by the baselines (especially DupPredictor and Doc2vec) seem not substantial, which may be explained by the fact that the positive cases (duplicate post pairs) and the negative cases (non-duplicate post pairs) are imbalanced (only 10\% pairs are positive). The AUC of a classifier is equivalent to the probability that the classifier will rank a randomly chosen positive instance higher than a randomly chosen negative instance~\cite{fawcett2006introduction}; in our case, the AUC indicates the probability that an approach can give a higher score (similarity or classification probability) for a duplicate post pair than a non-duplicate pair, which might be a trivial task given that the vast majority of the posts are very different from a candidate post (i.e., easy to treat the vast majority as non-duplicate pairs), thus both our approach and the baselines can achieve relatively high AUC scores. We further evaluate our approach and DupPredictor using Top-N accuracy, shown in Table~\ref{tab:results_topn}.

\textbf{Our approach outperforms the baseline in ranking the duplicate posts of a given post at the top.}
% Table~\ref{tab:results_topn} shows that our approach achieves a Top-1, 3, 5, 10, 30 accuracy scores of 21.2\% to 50.0\%, 35.3\% to 72.1\%, 44.2\% to 76.0\%, 55.1\% to 83.7\%, 74.6\% to 92.4\%,\heng{fill the numbers} respectively, which are 84.3\% to Y1\%, X3\% to Y3\%.... higher than that of the baseline (i.e., the DupPredictor approach).
% \xingfang{I think using *\% higher is confusing here, as the accuracies are actually already in \%. *\% higher can be either understood as x+*\% or x(1+*)\%.}\heng{Use either absolute or relative difference is fine. You can use ``absolute/relative difference'' wording to make it unambiguous.}
Table~\ref{tab:results_topn} shows that our approach achieves Top-30 accuracy of 74.6\% to 92.4\%, which are 22.8\% to 35.5\% higher than that of the baseline (i.e., the DupPredictor approach). It is worth noting that the variation in Top-N accuracy across different topics possibly stemming from\accepted{\heng{may be?}} the differences in dataset sizes, which impact the number of candidates. The superior performance of our method in terms of Top-N accuracy stands for better retrieval outcomes compared with the baseline approach, although the difference in their AUC scores is not very significant.

% \xingfang{I think listing all the top accuracies that have already been shown in the table is not necessary. I think just show the Top-30 accuracy as an example is enough. Listing too many numbers may make this paragraph unreadable.}\heng{sounds good}
\accepted{\heng{Explain the reason of the difference and the significance of the results (how our approach can be used).}
\heng{Also explain why the results are better form some topics and worse for others}}
%We further present the performance of DupPredictor and our approach with Top-N accuracy in Table~\ref{tab:results_topn}.
%\xingfang{is it necessary to show the top-n results for RQ1? We can not get the top-n for Doc2vec, as the results of my implementation are remarkablely poor than the results reported.}

% \usepackage{multirow}

\begin{table}
\centering
\caption{Top-N Accuracy Comparison: DupPredictor vs. Our Approach on CQADupStack}
% \vspace{-3mm}
\label{tab:results_topn}
\resizebox{\linewidth}{!}{
\begin{tabular}{l|llllll} 
\hline
Topic                    & Top-N    & 1      & 3      & 5      & 10     & 30      \\ 
\hline
\multirow{2}{*}{android} & Ours     & 42.4\% & 70.4\% & 76.0\% & 83.7\% & 92.4\%  \\
                         & DupPred. & 22.6\% & 42.2\% & 47.2\% & 52.3\% & 63.3\%  \\ 
\hline
\multirow{2}{*}{gis}     & Ours     & 50.0\% & 72.1\% & 75.0\% & 80.7\% & 91.4\%  \\
                         & DupPred. & 34.8\% & 48.2\% & 52.5\% & 58.9\% & 68.1\%  \\ 
\hline
\multirow{2}{*}{math.}   & Ours     & 25.4\% & 42.0\% & 50.0\% & 59.4\% & 74.6\%  \\
                         & DupPred. & 8.9\%  & 20.5\% & 26.0\% & 32.9\% & 43.2\%  \\ 
\hline
\multirow{2}{*}{prog.}   & Ours     & 23.6\% & 47.8\% & 54.5\% & 63.5\% & 80.9\%  \\
                         & DupPred. & 15.1\% & 24.3\% & 29.2\% & 36.2\% & 49.7\%  \\ 
\hline
\multirow{2}{*}{stats}   & Ours     & 45.1\% & 59.8\% & 67.7\% & 75.5\% & 90.2\%  \\
                         & DupPred. & 30.4\% & 36.3\% & 41.2\% & 49.0\% & 58.8\%  \\ 
\hline
\multirow{2}{*}{tex}     & Ours     & 29.5\% & 49.4\% & 56.1\% & 65.8\% & 78.7\%  \\
                         & DupPred. & 15.4\% & 23.8\% & 27.5\% & 33.0\% & 47.0\%  \\ 
\hline
\multirow{2}{*}{unix}    & Ours     & 37.5\% & 54.7\% & 63.5\% & 76.6\% & 88.0\%  \\
                         & DupPred. & 24.4\% & 39.4\% & 44.6\% & 51.3\% & 60.1\%  \\ 
\hline
\multirow{2}{*}{webm.}   & Ours     & 21.2\% & 35.3\% & 44.2\% & 55.1\% & 75.0\%  \\
                         & DupPred. & 11.5\% & 20.4\% & 22.3\% & 28.7\% & 39.5\%  \\ 
\hline
\multirow{2}{*}{wordp.}  & Ours     & 34.1\% & 52.3\% & 58.0\% & 75.0\% & 90.9\%  \\
                         & DupPred. & 22.0\% & 40.7\% & 50.5\% & 56.0\% & 68.1\%  \\
\hline
\end{tabular}
}
% \vspace{-3mm}
\end{table}

% \begin{table}
% \centering
% \label{tab:results_bm}
% \begin{tabular}{|l|c|c|c|c|c|c|c|c|c|} 
% \hline
% \multirow{2}{*}{Method}        & \multicolumn{9}{c|}{topics}                                                                                                                            \\ 
% \cline{2-10}
%                                & android        & gis            & mathematica    & programmers    & stats          & tex            & unix           & webmasters     & wordpress       \\ 
% \hline
% \multicolumn{1}{|c|}{GPT-3}    & \textbf{0.996} & \textbf{0.996} & \textbf{0.964} & \textbf{0.982} & 0.992          & \textbf{0.985} & \textbf{0.996} & \textbf{0.989} & \textbf{0.992}  \\ 
% \hline
% \multicolumn{1}{|c|}{Baseline} & 0.970          & 0.970          & 0.960          & 0.930          & \textbf{1.000} & 0.940          & 0.980          & 0.920          & 0.970           \\
% \hline
% \end{tabular}
% \end{table}

% \secdescription{comparison with benchmark}

\begin{answer*}{to RQ1}{}
Compared with the baseline approaches, our approach demonstrates notably enhanced performance in duplicate detection on the benchmark dataset. Furthermore, the outcomes suggest encouraging indications that its performance is more stable across technical topics.
\end{answer*}

% \textbf{RQ2: Can the performance be further improved through further supervised training? Which training configurations offer better potential results?}

% \textbf{RQ2: How do the Siamese structure and loss functions impact the performance of duplicate detection?}

\textbf{RQ2: How does our approach perform on our dataset constructed with all Stack Overflow duplicates? How do training settings (e.g., loss function, batch size) influence the performance?}

\setcounter{paragraph}{0}
\paragraph{\textbf{Objective}}
% In RQ1, we evaluate and compare the performance of our proposed approach in detecting duplicate technical posts with baseline approaches. 
In this research question, we aim to assess the effectiveness of our proposed approach in detecting duplicate posts in our constructed dataset (see Section~\ref{sec:data_construction}). In particular, we examine the efficacy of the Siamese structure and the supervised training process in refining the original GPT-3 embeddings. %Specifically, we look into whether the performance is improved through the training process that leverages the duplicate annotations in our constructed dataset and 
Besides, we investigate the impacts of the different training settings (i.e., loss functions, batch size) for the proposed Siamese structure\accepted{\heng{network or networks?}}.

% Based on the pre-trained GPT-3 embedding of technical posts, we aim to acquire a latent embedding which can better measure the duplicate relationship between the posts.

% In order to achieve optimal performance, we employ different training settings (i.e., loss functions, batch sizes) for the proposed Siamese-based networks. In this research question, we also aim to evaluate the performance under different training settings.

\paragraph{\textbf{Approach}}
In this research question, different from the experiments in RQ1, we conduct the experiments with our constructed dataset (see Section~\ref{sec:data_construction}). We split the dataset into train and test sets with an 80\%/20\% ratio. We respectively train the Siamese-based structure with two loss functions (i.e., triplet loss, MNR loss) introduced in Section~\ref{sec:loss}\accepted{\heng{wrong section ID}}, and evaluate the performance. We rank the candidate posts based on their similarity with the target posts, and report the Top-N scores of the results. We choose values of N as 1, 3, 5, 10, and 30. It is worth noting that since the number of candidate posts in our constructed dataset differs from that in the CQADupStack, direct comparison of the Top-N scores in this RQ and RQ1 cannot reflect the performance differences of the approaches.
% We use the ranking formulation for the detection task (i.e., for a target post, rank the candidate posts based on their similarity with the target post), and thus report the Top-N scores of the results. We choose values of N as 1, 3, 5, 10, and 30.
% The Top-N scores reflect the approach's ability to rank the duplicate candidates\heng{is this sentence redundant as top-N accuracy has been introduced previously?}. 
The default batch size is set to 64 for training with both the Triplet loss and MNR loss. Moreover, as the batch size serves as a hyperparameter for the Multiple Negative Ranking (MNR) Loss, we evaluate the performance under different training batch sizes. Our selection for the batch size spans from a relatively small value of 8 to a larger number of 256.

\paragraph{\textbf{Results and Analysis}}

\begin{table}[!h]
\centering
\caption{Top-N accuracy with different training settings.}
\label{tab:two_lossfn}
\begin{tabular}{c|ccc} 
\hline
\multicolumn{1}{l|}{\multirow{2}{*}{Top-N}} & \multirow{2}{*}{\begin{tabular}[c]{@{}c@{}}Original GPT-3\\embeddings\end{tabular}} & \multicolumn{2}{c}{\begin{tabular}[c]{@{}c@{}}Embeddings refined \\with Siamese structure\end{tabular}}  \\ 
\cline{3-4}
\multicolumn{1}{l|}{}                       &                                                                                     & Triplet Loss & MNR Loss                                                                                  \\ 
\hline
1                                           & 9.5\%                                                                               & 9.5\%        & 23.1\%                                                                                    \\
3                                           & 15.0\%                                                                              & 16.6\%       & 37.0\%                                                                                    \\
5                                           & 18.1\%                                                                              & 20.7\%       & 43.9\%                                                                                    \\
10                                          & 22.7\%                                                                              & 27.3\%       & 53.6\%                                                                                    \\
30                                          & 31.5\%                                                                              & 39.7\%       & 68.9\%                                                                                    \\
\hline
\end{tabular}
% \vspace{-1mm}
\end{table}

\textbf{Using a Siamese structure to refine the GPT-3 embeddings significantly improves the performance of duplicate post detection, especially when using the MNR loss function.}
Table~\ref{tab:two_lossfn} shows the Top-N accuracies of our Siamese-based approach separately trained with two studied loss functions. From the table, we can tell that the performance gains are evident after undergoing training with annotated duplicate data. Before training, the pre-trained GPT-3 embedding achieves a Top-30 score of 31.5\%. However, this score undergoes a substantial increase to 39.7\% and 68.9\%,\accepted{\heng{the numbers are wrong?}} respectively, subsequent to training with two distinct loss functions. This suggests an improvement in the ability of the obtained latent embeddings to represent the duplicate relationships among technical posts more accurately.

In terms of the selection of loss functions, the experimental results show that MNR loss outperforms the triplet loss by a considerable margin. The superior performance of the MNR loss may be attributed to its incorporation of multiple negative samples. Unlike the triplet loss, which only takes into account one negative sample for each training triplet, MNR loss treats all the candidates as negative samples in a batch and tries to maximize the distance between the anchors and negative samples. This characteristic enables better contrastive embedding learning by capturing a more comprehensive range of potential variations and similarities within the training data.

\textbf{Using a larger batch size further improves the performance of the Siamese structure with the MNR loss.}
Table \ref{tab:batch_size} illustrates the performance variation as we alter the batch size while training the model using the MNR loss. From the table, we can observe a slight but steady improvement in performance for all Top-N accuracies. When enlarging the batch size from 8 to 256, the Top-30 accuracy increases by 5.9\%\accepted{\heng{use relative improvement}\xingfang{Same as the previous comment.}\heng{Use either absolute or relative difference is fine. You can use ``absolute/relative difference'' wording to make it unambiguous.}}. This implies that having a greater number of samples serving as contrastive negatives in each batch, to some extent, can enhance the performance\accepted{\heng{this sentence seems grammatically wrong... I am not sure what it means}}. In particular, the improvement is more pronounced as the batch size is shifted from 8 to 16 for all Top-N accuracies. This implies further increasing the batch size may yield limited returns in terms of accuracy improvement.

\begin{table}[!t]
\centering
\caption{Top-N Accuracy of our approach with the MNR Loss Function using Different Batch Sizes}
% \vspace{-2mm}
\label{tab:batch_size}
\resizebox{\linewidth}{!}{
\begin{tabular}{c|cccccc} 
\hline
\multirow{2}{*}{TOP-N} & \multicolumn{6}{c}{Batch Size}                       \\ 
\cline{2-7}
                       & 8      & 16     & 32     & 64     & 128    & 256     \\ 
\hline
1                      & 21.3\% & 22.5\% & 22.8\% & 23.1\% & 23.3\% & 23.8\%  \\
3                      & 34.1\% & 36.2\% & 36.6\% & 37.0\% & 37.6\% & 38.1\%  \\
5                      & 40.6\% & 43.3\% & 43.5\% & 43.9\% & 44.8\% & 45.4\%  \\
10                     & 49.8\% & 53.0\% & 53.4\% & 53.6\% & 54.9\% & 55.4\%  \\
30                     & 64.5\% & 68.1\% & 68.7\% & 68.9\% & 70.2\% & 70.4\%  \\
\hline
\end{tabular}
}
% \vspace{-3mm}
\end{table}

\vspace{3mm}
\begin{answer*}{to RQ2}{}
Compared to directly using the GPT-3 embeddings, the performance of duplicate post detection can be significantly enhanced when utilizing a Siamese structure trained with the duplicate annotation available from Stack Overflow, in particular, when an MNR loss function is used. 
%for the training process outperforms the triplet loss. 
Besides, employing a larger training batch size can further benefit the MNR loss function. %with MNR loss can yield advantages for the training process.
% the Siamese-based structure trained with 
\end{answer*}
\vspace{3mm}

\textbf{RQ3: How does the performance of the models trained on data specific to a particular topic compare to that of the models trained on general data?}

% \xingfang{In-topic, general topic data}
% RQ3: Are there disparities among different post topics regarding duplicates? Do performances vary across different technical topics of posts?

\setcounter{paragraph}{0}
\paragraph{\textbf{Objective}} RQ1 shows that our proposed approach exhibits more stable performances across different technical topics than the baselines. However, there may exist different characteristics in the posts of different technical topics that may influence the performance of our proposed approach. In this research question, we aim to investigate the impacts of training our model with data of various technical topics (i.e., using the SO posts related to specific tags), compared to training the approach using the general SO post data. %explore the potential performance discrepancies of our approach among different technical topics by training and evaluating our method with post data of different topics.

\paragraph{\textbf{Approach}} 
%\heng{Re-organize the approach with three sub-titles: General-topic model, in-topic model, and cross-topic model. For each training method, explicitly specify the training and testing data}\xingfang{we don't have cross-topic model. I restructured the paragraph.}
We have obtained the model trained with posts of all topics and evaluated the performance over the general dataset in RQ2. As the training data contains posts from all topics, we call the model \textbf{\textit{general-topic model}} in this research question. In order to observe impacts of varying training data, we trained separate models from scratch with post data of specific topics. With the \textbf{\textit{in-topic models}} trained with topic-specific data and the general-topic model, we compare the performance on the topic-specific test data. We observe if there exist significant discrepancies among the performances that can reflect the disparities of topics. In this research question, we selected the top five topics (i.e., javascript, python, java, php, c\#) that have the largest amounts of duplicate posts as our research objects.
% We also conduct an experiment where we mutate the topic- specific training and test data of two topics (i.e., javascript and python questions) to exhibit cross-topic disparity. 

\paragraph{\textbf{Results and Analysis}} 

\begin{table*}[!ht]
\centering
\caption{Performance of the models trained with in-topic and general-topic training data}
\label{tab:inoutdm_exp}
\resizebox{0.75\linewidth}{!}{
% \begin{tabular}{|c|l|l|l|l|l|} 
% \hline
% \multirow{2}{*}{Topic} & \multicolumn{5}{c|}{\begin{tabular}[c]{@{}c@{}}Top-N accuracy\\(In-topic / general)\end{tabular}}                                                \\ 
% \cline{2-6}
%                         & \multicolumn{1}{c|}{Top-1} & \multicolumn{1}{c|}{Top-3} & \multicolumn{1}{c|}{Top-5} & \multicolumn{1}{c|}{Top-10} & \multicolumn{1}{c|}{Top-30}  \\ 
% \hline
% javascript              & 0.201 / 0.198              & 0.344 / 0.331              & 0.419 / 0.405              & 0.528 / 0.507               & 0.702 / 0.677                \\ 
% \hline
% python                  & 0.219 / 0.207              & 0.373 / 0.345              & 0.453 / 0.417              & 0.566 / 0.522               & 0.732 / 0.689                \\ 
% \hline
% java                    & 0.267 / 0.252              & 0.439 / 0.407              & 0.525 / 0.484              & 0.634 / 0.591               & 0.786 / 0.748                \\ 
% \hline
% php                     & 0.223 / 0.215              & 0.385 / 0.363              & 0.471 / 0.438              & 0.586 / 0.550               & 0.752 / 0.717                \\ 
% \hline
% c\#                     & 0.344 / 0.320              & 0.535 / 0.497              & 0.621 / 0.573              & 0.726 / 0.671               & 0.850 / 0.806                \\
% \hline
% \end{tabular}
\begin{tabular}{|c|l|l|l|l|l|} 
\hline
\multirow{2}{*}{Topic} & \multicolumn{5}{c|}{\begin{tabular}[c]{@{}c@{}}Top-N accuracy\\(In-topic / General-topic)\end{tabular}}                                                 \\ 
\cline{2-6}
                       & \multicolumn{1}{c|}{Top-1} & \multicolumn{1}{c|}{Top-3} & \multicolumn{1}{c|}{Top-5} & \multicolumn{1}{c|}{Top-10} & \multicolumn{1}{c|}{Top-30}  \\ 
\hline
javascript             & 20.1\% / 19.8\%            & 34.4\% / 33.1\%            & 41.9\% / 40.5\%            & 52.8\% / 50.7\%             & 70.2\% / 67.7\%              \\ 
\hline
python                 & 21.9\% / 20.7\%            & 37.3\% / 34.5\%            & 45.3\% / 41.7\%            & 56.6\% / 52.2\%             & 73.2\% / 68.9\%              \\ 
\hline
java                   & 26.7\% / 25.2\%            & 43.9\% / 40.7\%            & 52.5\% / 48.4\%            & 63.4\% / 59.1\%             & 78.6\% / 74.8\%              \\ 
\hline
php                    & 22.3\% / 21.5\%            & 38.5\% / 36.3\%            & 47.1\% / 43.8\%            & 58.6\% / 55.0\%             & 75.2\% / 71.7\%              \\ 
\hline
c\#                    & 34.4\% / 32.0\%            & 53.5\% / 49.7\%            & 62.1\% / 57.3\%            & 72.6\% / 67.1\%             & 85.0\% / 80.6\%              \\
\hline
\end{tabular}
}
 % \vspace{-2mm}
\end{table*}

\textbf{The duplicate post detection models trained within a topic achieve better results than those trained with data from all topics.}
Table~\ref{tab:inoutdm_exp} shows the Top-N accuracies of in-topic model and general-topic model on in-topic test sets. We can observe a consistent performance gap between two types of models on the same test set. Compared with the model trained with data containing posts from all topics, models trained with corresponding single topics work significantly better (e.g., with a Top-30 accuracy improvement from 2.5\% to 4.3\% \accepted{\heng{X\% to Y\%}}), albeit having fewer training samples. The results suggest that there exist discrepancies in the characteristics of duplicate posts of different topics. %\heng{Would be good to explain the difference using conjecture or an example}

The better performance achieved by the within-topic models may be explained by the fact that most of the duplicates share topics (i.e., tags) (our analysis in Section~\ref{sec:preliminary}\accepted{\heng{refer to the section}} shows that 95.5\%\accepted{\heng{X\%}} of the duplicate posts have at least one common tag between the duplicate pair). Therefore, it is recommended to use topics (tags) as a filter to detect duplicate posts.

\begin{answer*}{to RQ3}{}
Our approach achieves a better performance when trained within a specific topic than when trained using general data. % or the data of a different topic.
%There exist discrepancies in performance when models are trained with posts of different topics, which 
The results reflect disparities in the characteristics of posts from different technical topics regarding duplication.
This factor should be considered when applying the approach in practical applications.
% Future work on duplicate post detection should consider this factor and train separate models for different topics if necessary.
% consider training the models within a technical topic as it is not only more efficient but also more accurate.
\end{answer*}

% \textbf{RQX: Summarization (case study)/ explainability?}

%\section{Discussion: How is the ability of the model to find the duplicate relation of unlabelled posts?} 
\section{Discussion: Can our approach detect unmarked duplicate posts on Stack Overflow?} 
\label{sec:discussion}
% \xingfang{Annotations by Qiaolin \& kaveh: \url{https://docs.google.com/spreadsheets/d/1h3KzbIKp32b4QQ5Kmb_MZ4O4yYwtCTO7/edit?usp=sharing&ouid=102449022365724027511&rtpof=true&sd=true}}

% \setcounter{paragraph}{0}
% \xingfang{Do we need to have the 'related' in our labelling? 'related' or not is very objective.}\heng{Not necessary I think}
In this session, we aim to explore the potential of our proposed approach in finding unlabeled duplicate posts. Specifically, our goal is to see if the ranking results predicted by our proposed method can actually help in finding unlabelled duplicate posts besides those that have already been annotated.

To achieve our goal, we conduct a manual study on ten randomly chosen queries. We manually examine the titles and descriptions of the queries and their top-10 candidates to check if there exist any unlabelled duplicate posts. We annotate the candidates into three categories: ground truth, duplicate and non-duplicate. Ground truths are the ones marked as duplicates by the SO. The duplicate candidates are posts that deemed to be duplicates of the original posts by coders. The others are considered to be non-duplicates.

% To achieve our goal, we conduct a manual study on the ranking results of ten randomly chosen queries, which involves 115 posts \heng{not clear why 115 pots. need an explanation (e.g., how many posts are examined for each query/target post)}. We manually examine the titles and descriptions of the top-10 candidates of each query to check if there exist any unlabelled duplicate posts. According to their similarity to the original query, we simply annotate the candidates into three categories duplicate, related and unrelated\heng{update}. If the ground truth appears in the candidate list, we simply label it as ground truth. The candidates marked as duplicate are posts that deemed to be the duplicate of the original post. The related candidates are posts authors consider to be different questions from the queries but closely related to the topic and techniques discussed in the original posts (e.g., reverse questions). The candidates fall into other situation are considered to be unrelated.\heng{update}

% \xingfang{there are 92 candidates in total, we excludes 8 that are GT from the 100 candidates. A\&B agreed on 9 candidates as duplicates. A marked 2 as duplicate which B thought to be Non-dup, and vice versa. -> Cohen's kappa tobe 0.327.}\heng{it seems the kappa value should be 0.793? $p0=(9+79)/92; p_y = (11/92)*(11/92); p_n=(81/92)*(81/92); pe=p_y + p_n; kappa = (p0-pe)/(1-pe) = 0.793$}

This manual labeling process, which involves 92 candidate posts (8 of the 100 candidates are ground truth posts, and thus do not require further annotation), includes the participation of three doctoral students majoring in computer engineering. 
% The total amount of candidates to be annotated is 100. However, 115 posts were checked in this manual labeling process, including the original posts, their ground truth posts and top-ranking candidates. 
In the first round, two of them independently annotated the posts, and reached a Cohen's kappa of 0.793, which indicates a \textit{substantial} agreement. In the second round, the third coder checked the disagreements from the first round, and resolved all discrepancies through discussion. Hence, our labeling outcomes can be considered reliable.

We take the example mentioned in Section~\ref{sec:background} as an illustration of the annotation results of the manual study. Table~\ref{tab:case_study} shows the sampled query and the ranking results of the duplicate candidates generated with our approach. According to our manual examination, five of the candidates are considered duplicates of the original post, although their contexts may differ from those of the query post to some extent, while SO only marked one (i.e., the ground truth). For example, Candidate 1 describes in detail the use case while the core of the question remains to be using JavaScript to implement the copy function. Candidate 6 has a distinct need for texts that need to be copied to be rich ones.

\textbf{Our approach can assist in finding duplicate posts that are not labeled by SO.}
From the manual labeling, we found that seven out of the ten posts have unlabeled duplicates. Among all the 100 candidates of the query posts, our approach found 21 duplicate posts, among which 13 were unlabelled ones. The outcome suggests that our approach has the ability to discover unlabeled duplicates on Stack Overflow data, although it still requires additional human efforts to examine the ranking results.

\begin{table}
\centering
\caption{A Sample Query and Its Top-10 Candidates}
% \vspace{-3mm}
\label{tab:case_study}
\resizebox{\linewidth}{!}{
\begin{tabular}{|l|l|p{0.5\columnwidth}|c|} 
\hline
Post         & ID & Title                                                                                                                  & Annotation       \\ 
\hline
Query        & 34954370    & Copy sentence to Clipboard using simple JS                                                                             & -                \\ 
\hline
Candidate 1  & 43638872    & Copy string to clipboard initiated by click on injected element in JavaScript                                          & Duplicate        \\ 
\hline
Candidate 2  & 400212      & How do I copy to the clipboard in JavaScript?                                                                          & Ground Truth     \\ 
\hline
Candidate 3  & 2176861     & JavaScript get clipboard data on paste event (Cross browser)                                                           & Non-duplicate  \\ 
\hline
Candidate 4  & 23475342    & Get clipboard contents with Greasemonkey                                                                               & Non-duplicate  \\ 
\hline
Candidate 5  & 6300213     & Copy selected text to the clipboard WITHOUT using flash - must be cross-browser                                        & Duplicate        \\ 
\hline
Candidate 6  & 23934656    & How can I copy rich text contents to the clipboard with JavaScript?                                                    & Duplicate        \\ 
\hline
Candidate 7  & 70147931    & Insert a link into selected text using JS (losing window.getSelection() value when user focuses on input to enter URL) & Non-duplicate          \\ 
\hline
Candidate 8  & 16748735    & Clear clipboard to prohibit unauthorised copying, insert message?                                                      & Non-duplicate          \\ 
\hline
Candidate 9  & 36639681    & How to copy text from a div to clipboard                                                                               & Duplicate        \\ 
\hline
Candidate 10 & 50633601    & Is it possible to paste from clipboard onclick in Javascript?                                                          & Non-duplicate\\
\hline
\end{tabular}
}
% \vspace{-3mm}
\end{table}

\section{Threats to Validity} \label{sec:threats}

\noindent \textbf{Construct Validity.} 
There exist posts that are not yet identified as duplicates (i.e., false negatives) in our dataset for training and evaluating our model. It is possible that the negative training samples can actually be positive ones, which influence the performance of our model. Nevertheless, we employed a random sampling strategy, and given the large size of the training dataset, the likelihood of encountering a false negative in the training sample is very low, which helps mitigate the threat. Future studies can refine the dataset by mining unlabelled duplicates with our proposed detection methods and manually checking the potential duplicates, and thus, false negative samples can be reduced.

\noindent \textbf{Internal Validity.} 
Our training settings for our proposed model may not be optimal, which may influence the accuracy of our evaluations. As the number of hyper-parameters and variables (e.g., training rate, batch size, etc.) in the training process is large, we were not able to try all the combinations to get optimal results. We referred to previous works to mitigate this threat and chose the common settings to enable a more fair evaluation. Due to constraints in computational resources and time allocations, the Siamese structure employed in our experiments might not be optimal. These limitations impeded a thorough exploration of ideal structural settings. Despite these challenges, we attempted various adjustments, including altering the number of hidden layers, incorporating batch normalization techniques, and so forth. However, these modifications did not lead to notable performance improvements. In our proposed approach, fields (e.g., tag and code block), which may contain valuable and informative for duplicate detection, are omitted, which may cause information loss and result in suboptimal results. Future works may consider taking in these fields to achieve a better result. Moreover, the manual study and annotation involved in this study may suffer from subjectivity and even the authors' bias. The involvement of three experienced persons reduces the bias.

\noindent \textbf{External Validity.}
The datasets used and constructed in this study are limited to specific forums (e.g., Stack Overflow).
% There are many other technical forums in the domain of software engineering. 
Therefore, our results may not be applied to technical posts from other forums. However, Stack Overflow is one of the most popular technical forums which covers relatively more diverse topics. %Besides the large number of posts it contains, the topics and technical domains covered by Stack Overflow are relatively more diverse.
Besides, the benchmark dataset used in our evaluation in RQ1 also contains data from other Stack Exchange forums.
% To further mitigate this threat of lack of generalizability, we include all the duplicate pairs available when constructing the dataset. 
Nevertheless, further study can evaluate our approach on data from other technical forums.

\section{Conclusions} \label{sec:conclusions}

% In this research, we explore utilizing and refining the GPT-3 embeddings to address the challenge of identifying duplicate questions in technical forum posts. Besides adopting a public available benchmark dataset, we constructed a more up-to-date duplicate dataset from a recent Stack Overflow dump. With these two dataset, our evaluation evaluate and compare the performance of our approach with baseline approaches and explore the characteristics of our method. The evaluation shows that the refined embeddings exhibit stronger and more stable ability to capture duplicate relations among technical posts. Furthermore, we conduct a manual study to gain more precise comprehension of its capacity in finding unlabeled duplicates. Future works could incorporate more meta-information of the technical posts to the proposed approach and conduct more in-depth investigations on the discrepancies among different technical domains.

In this research, we leverage GPT-3 embeddings and a Siamese-based structure to tackle the challenge of identifying duplicate questions in technical forum posts. 
% We utilize both a publicly accessible benchmark dataset and a newly constructed duplicate dataset extracted from a recent Stack Overflow dump to conduct a comprehensive evaluation of our approach. 
Our evaluation assesses the performance of our approach through a comparison with baseline methods and delves into the characteristics of our methodology. The results demonstrate that our approach is accurate and robust at detecting duplicate relations among technical posts. Additionally, a manual study of the detection results of our approach reveals that it can help identify missed labels of duplicate posts.
Our study highlights the issues present in the existing management mechanism for duplicate posts on technical forums such as Stack Overflow.
Technical forum providers may leverage our proposed idea, approach, and shared implementation to augment their manual effort for duplicate post detection (e.g., combining automated recommendations and domain expertises).
%our manual study help us gain deeper insights into its capacity to identify unlabeled duplicates. 
Future research could improve our approach by leveraging the fast-growing capacity of LLMs or incorporating more meta-information of the technical posts into the proposed approach. %, or build on our study and conduct more in-depth investigations on the discrepancies among different technical domains.

\section*{Acknowledgement}
The authors express their gratitude to the anonymous reviewers for their valuable and constructive comments. Additionally, sincere thanks are extended to Shunsuke Mori (Waseda Univ.), Kaveh Shahedi and Qiaolin Qin (PolyMTL) for their invaluable suggestions and assistance throughout this work. We would like to gratefully acknowledge the Natural Sciences and Engineering Research Council of Canada (NSERC), Mitacs, and Peritus.ai for funding this project.

% \begin{thebibliography}{1}

% \end{thebibliography}
\balance
\bibliography{duplicate}

\begin{thebibliography}{10}

\bibitem{abric2019can}
Durham Abric, Oliver~E Clark, Matthew Caminiti, Keheliya Gallaba, and Shane McIntosh.
\newblock Can duplicate questions on stack overflow benefit the software development community?
\newblock In {\em 2019 IEEE/ACM 16th International Conference on Mining Software Repositories (MSR)}, pages 230--234. IEEE, 2019.

\bibitem{correa2013fit}
Denzil Correa and Ashish Sureka.
\newblock Fit or unfit: analysis and prediction of'closed questions' on stack overflow.
\newblock In {\em Proceedings of the first ACM conference on Online social networks}, pages 201--212, 2013.

\bibitem{so_duplicate}
Jeff Atwood.
\newblock Dr. strangedupe: Or, how i learned to stop worrying and love duplication.
\newblock \url{https://stackoverflow.blog/2010/11/16/dr-strangedupe-or-how-i-learned-to-stop-worrying-and-love-duplication}.

\bibitem{zhang2015multi}
Yun Zhang, David Lo, Xin Xia, and Jian-Ling Sun.
\newblock Multi-factor duplicate question detection in stack overflow.
\newblock {\em Journal of Computer Science and Technology}, 30:981--997, 2015.

\bibitem{ahasanuzzaman2016mining}
Muhammad Ahasanuzzaman, Muhammad Asaduzzaman, Chanchal~K Roy, and Kevin~A Schneider.
\newblock Mining duplicate questions in stack overflow.
\newblock In {\em Proceedings of the 13th International Conference on Mining Software Repositories}, pages 402--412, 2016.

\bibitem{lau2016empirical}
Jey~Han Lau and Timothy Baldwin.
\newblock An empirical evaluation of doc2vec with practical insights into document embedding generation.
\newblock {\em arXiv preprint arXiv:1607.05368}, 2016.

\bibitem{zhang2017feature}
Wei~Emma Zhang, Quan~Z Sheng, Yanjun Shu, and Vanh~Khuyen Nguyen.
\newblock Feature analysis for duplicate detection in programming qa communities.
\newblock In {\em Advanced Data Mining and Applications: 13th International Conference, ADMA 2017, Singapore, November 5--6, 2017, Proceedings 13}, pages 623--638. Springer, 2017.

\bibitem{zhang2017detecting}
Wei~Emma Zhang, Quan~Z Sheng, Jey~Han Lau, and Ermyas Abebe.
\newblock Detecting duplicate posts in programming qa communities via latent semantics and association rules.
\newblock In {\em Proceedings of the 26th International Conference on World Wide Web}, pages 1221--1229, 2017.

\bibitem{zhang2018duplicate}
Wei~Emma Zhang, Quan~Z Sheng, Jey~Han Lau, Ermyas Abebe, and Wenjie Ruan.
\newblock Duplicate detection in programming question answering communities.
\newblock {\em ACM Transactions on Internet Technology (TOIT)}, 18(3):1--21, 2018.

\bibitem{silva2018duplicate}
Rodrigo~FG Silva, Kl{\'e}risson Paix{\~a}o, and Marcelo de~Almeida~Maia.
\newblock Duplicate question detection in stack overflow: A reproducibility study.
\newblock In {\em 2018 IEEE 25th international conference on software analysis, evolution and reengineering (SANER)}, pages 572--581. IEEE, 2018.

\bibitem{wang2020duplicate}
Liting Wang, Li~Zhang, and Jing Jiang.
\newblock Duplicate question detection with deep learning in stack overflow.
\newblock {\em IEEE Access}, 8:25964--25975, 2020.

\bibitem{koch2015siamese}
Gregory Koch.
\newblock {\em Siamese Neural Networks for One-Shot Image Recognition}.
\newblock PhD thesis, University of Toronto, 2015.

\bibitem{kamienski2023analyzing}
Arthur Kamienski, Abram Hindle, and Cor-Paul Bezemer.
\newblock Analyzing techniques for duplicate question detection on q\&a websites for game developers.
\newblock {\em Empirical Software Engineering}, 28(1):17, 2023.

\bibitem{bengio2000neural}
Yoshua Bengio, R{\'e}jean Ducharme, and Pascal Vincent.
\newblock A neural probabilistic language model.
\newblock {\em Advances in neural information processing systems}, 13, 2000.

\bibitem{mikolov2013distributed}
Tomas Mikolov, Ilya Sutskever, Kai Chen, Greg~S Corrado, and Jeff Dean.
\newblock Distributed representations of words and phrases and their compositionality.
\newblock {\em Advances in neural information processing systems}, 26, 2013.

\bibitem{pennington2014glove}
Jeffrey Pennington, Richard Socher, and Christopher~D Manning.
\newblock Glove: Global vectors for word representation.
\newblock In {\em Proceedings of the 2014 conference on empirical methods in natural language processing (EMNLP)}, pages 1532--1543, 2014.

\bibitem{le2014distributed}
Quoc Le and Tomas Mikolov.
\newblock Distributed representations of sentences and documents.
\newblock In {\em International conference on machine learning}, pages 1188--1196. PMLR, 2014.

\bibitem{li2014neural}
Peng Li, Yang Liu, Maosong Sun, Tatsuya Izuha, and Dakun Zhang.
\newblock A neural reordering model for phrase-based translation.
\newblock In {\em Proceedings of COLING 2014, the 25th International Conference on Computational Linguistics: Technical Papers}, pages 1897--1907, 2014.

\bibitem{zhang2014bilingually}
Jiajun Zhang, Shujie Liu, Mu~Li, Ming Zhou, and Chengqing Zong.
\newblock Bilingually-constrained phrase embeddings for machine translation.
\newblock In {\em Proceedings of the 52nd Annual Meeting of the Association for Computational Linguistics (Volume 1: Long Papers)}, pages 111--121, 2014.

\bibitem{devlin2018bert}
Jacob Devlin, Ming-Wei Chang, Kenton Lee, and Kristina Toutanova.
\newblock Bert: Pre-training of deep bidirectional transformers for language understanding.
\newblock {\em arXiv preprint arXiv:1810.04805}, 2018.

\bibitem{brown2020language}
Tom Brown, Benjamin Mann, Nick Ryder, Melanie Subbiah, Jared~D Kaplan, Prafulla Dhariwal, Arvind Neelakantan, Pranav Shyam, Girish Sastry, Amanda Askell, et~al.
\newblock Language models are few-shot learners.
\newblock {\em Advances in neural information processing systems}, 33:1877--1901, 2020.

\bibitem{wang2020measurement}
Jiapeng Wang and Yihong Dong.
\newblock Measurement of text similarity: a survey.
\newblock {\em Information}, 11(9):421, 2020.

\bibitem{xu2021post2vec}
Bowen Xu, Thong Hoang, Abhishek Sharma, Chengran Yang, Xin Xia, and David Lo.
\newblock Post2vec: Learning distributed representations of stack overflow posts.
\newblock {\em IEEE Transactions on Software Engineering}, 48(9):3423--3441, 2021.

\bibitem{kucuk2021characterizing}
Berfin Kucuk and Eray Tuzun.
\newblock Characterizing duplicate bugs: An empirical analysis.
\newblock In {\em 2021 IEEE International Conference on Software Analysis, Evolution and Reengineering (SANER)}, pages 661--668. IEEE, 2021.

\bibitem{sun2011towards}
Chengnian Sun, David Lo, Siau-Cheng Khoo, and Jing Jiang.
\newblock Towards more accurate retrieval of duplicate bug reports.
\newblock In {\em 2011 26th IEEE/ACM International Conference on Automated Software Engineering (ASE 2011)}, pages 253--262. IEEE, 2011.

\bibitem{zhang2023duplicate}
Ting Zhang, DongGyun Han, Venkatesh Vinayakarao, Ivana~Clairine Irsan, Bowen Xu, Ferdian Thung, David Lo, and Lingxiao Jiang.
\newblock Duplicate bug report detection: How far are we?
\newblock {\em ACM Transactions on Software Engineering and Methodology}, 32(4):1--32, 2023.

\bibitem{robertson2004simple}
Stephen Robertson, Hugo Zaragoza, and Michael Taylor.
\newblock Simple bm25 extension to multiple weighted fields.
\newblock In {\em Proceedings of the thirteenth ACM international conference on Information and knowledge management}, pages 42--49, 2004.

\bibitem{deshmukh2017towards}
Jayati Deshmukh, KM~Annervaz, Sanjay Podder, Shubhashis Sengupta, and Neville Dubash.
\newblock Towards accurate duplicate bug retrieval using deep learning techniques.
\newblock In {\em 2017 IEEE International conference on software maintenance and evolution (ICSME)}, pages 115--124. IEEE, 2017.

\bibitem{he2020duplicate}
Jianjun He, Ling Xu, Meng Yan, Xin Xia, and Yan Lei.
\newblock Duplicate bug report detection using dual-channel convolutional neural networks.
\newblock In {\em Proceedings of the 28th International Conference on Program Comprehension}, pages 117--127, 2020.

\bibitem{wang2019detecting}
Liting Wang, Li~Zhang, and Jing Jiang.
\newblock Detecting duplicate questions in stack overflow via deep learning approaches.
\newblock In {\em 2019 26th Asia-Pacific Software Engineering Conference (APSEC)}, pages 506--513. IEEE, 2019.

\bibitem{bromley1993signature}
Jane Bromley, Isabelle Guyon, Yann LeCun, Eduard S{\"a}ckinger, and Roopak Shah.
\newblock Signature verification using a" siamese" time delay neural network.
\newblock {\em Advances in neural information processing systems}, 6, 1993.

\bibitem{hoffer2015deep}
Elad Hoffer and Nir Ailon.
\newblock Deep metric learning using triplet network.
\newblock In {\em Similarity-Based Pattern Recognition: Third International Workshop, SIMBAD 2015, Copenhagen, Denmark, October 12-14, 2015. Proceedings 3}, pages 84--92. Springer, 2015.

\bibitem{henderson2017efficient}
Matthew Henderson, Rami Al-Rfou, Brian Strope, Yun-Hsuan Sung, L{\'a}szl{\'o} Luk{\'a}cs, Ruiqi Guo, Sanjiv Kumar, Balint Miklos, and Ray Kurzweil.
\newblock Efficient natural language response suggestion for smart reply.
\newblock {\em arXiv preprint arXiv:1705.00652}, 2017.

\bibitem{hoogeveen2015cqadupstack}
Doris Hoogeveen, Karin~M Verspoor, and Timothy Baldwin.
\newblock Cqadupstack: A benchmark data set for community question-answering research.
\newblock In {\em Proceedings of the 20th Australasian document computing symposium}, pages 1--8, 2015.

\bibitem{fawcett2006introduction}
Tom Fawcett.
\newblock An introduction to roc analysis.
\newblock {\em Pattern recognition letters}, 27(8):861--874, 2006.

\end{thebibliography}

\end{document}